# Optimal Strategies for Bidding Agents Participating in Simultaneous Vickrey Auctions with Perfect Substitutes

**Enrico H. Gerding**                                    EG@ECS.SOTON.AC.UK
**Rajdeep K. Dash**                                     RKD@ECS.SOTON.AC.UK
**Andrew Byde**                                        AB06V@ECS.SOTON.AC.UK
**Nicholas R. Jennings**                                NRJ@ECS.SOTON.AC.UK
*Intelligence, Agents, Multimedia Group*
*School of Electronics and Computer Science*
*University of Southampton, Southampton, UK*

## Abstract

We derive optimal strategies for a bidding agent that participates in multiple, simultaneous second-price auctions with perfect substitutes. We prove that, if everyone else bids locally in a single auction, the global bidder should always place non-zero bids in all available auctions, provided there are no budget constraints. With a budget, however, the optimal strategy is to bid locally if this budget is equal or less than the valuation. Furthermore, for a wide range of valuation distributions, we prove that the problem of finding the optimal bids reduces to two dimensions if all auctions are identical. Finally, we address markets with both sequential and simultaneous auctions, non-identical auctions, and the allocative efficiency of the market.

## 1. Introduction

In recent years, there has been a surge in the application of auctions, both online and within multi-agent systems (Wellman, Greenwald, & Stone, 2007; Clearwater, 1996; Gerding, Rogers, Dash, & Jennings, 2007b; Rogers, David, & Jennings, 2005; Rosenthal & Wang, 1996; Roth & Ockenfels, 2002; Dash, Parkes, & Jennings, 2003). As a result, there are an increasing number of auctions offering very similar or even identical goods and services.[1] To take advantage of this fact, automated support needs to be developed that can monitor, bid in, and make decisions across this large set of possibilities. Such software, in the form of intelligent bidding agents (hereafter shortened to bidder), can increase the likelihood of winning an item and at a lower price, and thus result in a considerable advantage for a buyer. Now, whereas participating in many auctions is an arduous task when done manually, such a problem is ideally suited for autonomous agents who can execute the proper actions on a buyer's behalf (Stone, Schapire, Littman, Csirik, & McAllester, 2003). To this end, in this paper we devise and analyse optimal bidding strategies for one such auction setting — namely, a bidder that participates in multiple, simultaneous second-price auctions for goods that are perfect substitutes. As we will show, however, this analysis also applies to a wider context with markets consisting of both sequential (i.e., where the auctions close one after the other), as well as simultaneous (i.e., where the auctions close at the same time) auctions.

---

1. In eBay alone, for example, there are often hundreds or sometimes even thousands of concurrent auctions running worldwide selling such substitutable items. To illustrate, at the time of writing, over 1600 eBay auctions were selling the Apple iPhone worldwide.

To date, much of the existing literature on multiple auctions focuses either on sequential auctions (Krishna, 2002) or on simultaneous auctions with complementarities, where the value of items together is greater than the sum of the individual items (see Section 2 for related research on simultaneous auctions). In contrast, here we consider bidding strategies for the case of simultaneous auctions and perfect substitutes. In particular, our focus is on Vickrey or second-price sealed bid auctions. We choose these because they have a low communication overhead (in terms of the number of required interactions) and are well known for their capacity to induce truthful bidding. As a result, this type of auction and its generalisations, such as the Vickrey-Clarke-Groves mechanism, have been used in a number of multi-agent system settings (Dash, Rogers, Reece, Roberts, & Jennings, 2005; Varian, 1995; Mes, van der Heijden, & van Harten, 2007; Dash, Vytelingum, Rogers, David, & Jennings, 2007). Moreover, these auctions are (weakly) strategically equivalent to the widely used English auctions.[2] However, we find that, when there are multiple such auctions running simultaneously, truthful bidding is no longer optimal. Given this, we characterise, for the first time, a bidding agent's utility-maximising strategy for bidding in any number of such auctions and for any type of bidder valuation distribution.

In more detail, we consider a market where a single bidder, called the *global bidder*, can bid in any number of auctions, whereas the other bidders, called the *local bidders*, are assumed to bid in only a single auction. We distinguish between two types of settings for this market: one in which the auctions are identical and the global bidder is indifferent between the auctions, and one in which the global bidder prefers some auctions over others. For these settings our main results are as follows:

- Whereas in the case of a single second-price auction a bidder has a weakly dominant strategy to bid its true value, this is no longer the case when there are several simultaneous auctions. The best strategy for a global bidder is then to bid below its true value.

- We prove that, even if a global bidder requires only one item and assuming free disposal, the expected utility is maximised by participating (i.e., bidding a non-zero amount) in all the auctions that are selling the desired item.

- Finding the optimal bid for each auction can be an arduous task when considering all possible combinations. However, when the global bidder is indifferent between the auctions, we are able to significantly reduce the search space for common bidder valuation distributions. As a result, optimal bids can be efficiently calculated for any number of auctions. Although the setting where the global bidder has preferences over auctions is more involved, we can still apply analytical methods to obtain tractable optimal results.

- We prove that, if the auctions are identical, the bidder's expected utility is maximised by bidding either uniformly across all auctions, or relatively high in one of the auctions, and the same, low value in all others (which of these two behaviours is optimal depends on the bidder valuation and market conditions such as the number of other bidders). If this is not the case and a global bidder has different preferences over the auctions, we show that it is optimal to bid relatively higher in auctions which are preferred over other auctions.

---

2. More specifically, Vickrey and English auction are strategically equivalent assuming *private valuations* of the good, i.e., where a bidder's value for the item remains unchanged during the bidding process.

- We argue that, even though a global bidder has a significantly higher expected utility than a local one, not all bidders should necessarily bid globally. For example, if bidders have budget considerations which constrain the amount they can bid, we show analytically that, if the budget is equal or less than the valuation, it is optimal to bid in a single auction under certain conditions.

Finally, we consider the issue of market efficiency when there are such simultaneous auctions. Efficiency is an important system-wide consideration within the area of multi-agent systems since it characterises how well the allocations in the system maximise the overall utility (Dash et al., 2003). Now, efficiency is maximised when the goods are allocated to those who value them the most. However, a certain amount of inefficiency is inherent to a distributed market where the auctions are held separately. Given this, in this paper, we measure the efficiency of markets with local bidders only and consider the impact of global bidders on this inefficiency. In so doing, we find that the presence of a global bidder generally has a positive impact on the efficiency.

The remainder of the paper is structured as follows. We first discuss related work in Section 2. In Section 3 we describe the bidders and the auctions in more detail. In Section 4 we characterise the optimal bidding behaviour for the base setting where all auctions are identical, and in Section 5 we explore a number of extensions: non-identical auctions, budget constraints and sequential auctions. In Section 6 we address the market efficiency and the impact of a global bidder. Finally, Section 7 concludes. Most of the proofs are placed in the appendix.

## 2. Related Work

Research in the area of simultaneous auctions can be segmented along two broad lines. On the one hand, there is the game-theoretic analysis of simultaneous auctions which concentrates on studying the equilibrium strategy of rational agents (Engelbrecht-Wiggans & Weber, 1979; Krishna & Rosenthal, 1996; Lang & Rosenthal, 1991; Rosenthal & Wang, 1996; Szentes & Rosenthal, 2003). Such analyses are typically used when the auction format employed in the simultaneous auctions is the same (e.g., there are $m$ Vickrey auctions or $m$ first-price auctions). On the other hand, heuristic strategies have been developed for more complex settings when the sellers offer different types of auctions or the buyers need to buy bundles of goods over distributed auctions (Stone et al., 2003; Byde, Preist, & Jennings, 2002; Yuen, Byde, & Jennings, 2006; Greenwald, Kirby, Reiter, & Boyan, 2001; Greenwald & Boyan, 2004; Wellman, Reeves, Lochner, & Vorobeychik, 2004). This paper adopts the former approach in studying a market of $m$ simultaneous Vickrey auctions since this approach yields provably optimal bidding strategies.

Related to our approach is the seminal paper by Engelbrecht-Wiggans and Weber (1979), which provides one of the starting points for the game-theoretic analysis of distributed markets where buyers have substitutable goods. Their work analyses a market consisting of couples having equal valuations that want to bid for a dresser. Thus, the couple's bid space can at most contain two bids since the husband and wife can be at most at two geographically distributed auctions simultaneously. They derive a mixed strategy Nash equilibrium for the special case where the number of buyers is large. Our analysis differs from theirs in that we use a decision-theoretic approach as opposed to a game-theoretic one. As argued among others by Rothkopf (2007), and Jiang and Leyton-Brown (2007), a decision-theoretic analysis is often sufficient in practice in the case of auctions.[3] Moreover, the decision-theoretic framework allows us to study a more complex setting in which bidders have different valuations and the global bidder can bid in all the auctions simultaneously (which is entirely possible for online auctions).

Subsequently, Krishna and Rosenthal (1996) studied the case of simultaneous auctions with complementary goods. They analyse the case of both local and global bidders and characterise the bidding of the buyers and resultant market efficiency. The setting provided by Krishna and Rosenthal (1996) is further extended to the case of common values by Rosenthal and Wang (1996). However, neither of these works extend easily to the case of substitutable goods which we consider. This case is studied by Szentes and Rosenthal (2003), but the scenario considered is restricted to three sellers and two global bidders and with each bidder having the same value (and thereby knowing the value of other bidders). The space of symmetric mixed equilibrium strategies is derived for this special case. However, as mentioned earlier, our results are based on decision theory, rather than game theory, but our setting is more general (i.e., we consider an arbitrary number of auctions). A number of other authors study settings where bidders face multiple simultaneous sealed-bid auctions, e.g. McAfee (1993), Peters and Severinov (1997), Gerding et al. (2007b), Leyton-Brown, Shoham, and Tennenholtz (2000). All of these papers, however, assume that bidders bid in only a single auction and choose this auction randomly. As we show here, however, this is not optimal. Finally, Shehory (2002) considers the case of concurrent *English* auctions, in which bidding algorithms are developed for buyers with different risk attitudes. However, in his setting the auctions never close at the same time, and he forces the bids to be the same across auctions. Although this strategy may be effective for the described setting, as we show in this paper, this is not always optimal when auctions close simultaneously (and the buyers bid very late in the auctions).

Related to this paper is also the literature that considers bidding with budget constraints. Although it is beyond the scope of this paper to provide a full literature review on this topic, here we highlight the most relevant work. A more extensive recent overview is presented by Pitchik (2006). A number of papers, such as those by Rothkopf (1977) and Palfrey (1980), study optimal bidding in simultaneous first-price auctions when bidders have constraints on *exposure*, which refers to the sum of bids. These papers, however, make the very strong assumption that the value accrued in one auction is independent of that in others. In other words, winning or losing one auction does not affect the expected utility in other auctions. In contrast, here we consider complete substitutes where bidders require only a single item. As a result, if one of the auctions is won, the value accrued in other auctions is zero. This interdependency between auctions makes the analysis sig-nificantly more difficult. Krishna and Benoit (2001) also consider this interdependency, but the valuations and budget constraints are assumed to be common knowledge (and the problem would thus be trivial without any budget constraints). Others, such as Che and Gale (1998), consider the effect of budget constraints in single auctions, where bidder types consist of two dimensions: their valuation and the budget. Sequential auctions have also been extensively studied with regards to budget constraints, e.g., by Krishna and Benoit (2001), Pitchik (2006). Furthermore, in addition to constraints on exposure, other types of budget constraints have been considered, in particular limits on the *expected* expenditures (Engelbrecht-Wiggans, 1987). None of these papers, however, address the particular setting that we consider here.

---

3. Essentially, a decision-theoretic setup requires only (1) information about the distribution of the *best competitive bid*, and (2) that the bidders optimize their decision given this assessment (Rothkopf, 2007). Such information about the distribution can be obtained by observing previous bids, e.g., using a learning approach as described by Jiang and Leyton-Brown (2007).

Finally, a number of researchers have investigated online auctions such as eBay. On the one hand, Hendricks, Onur, and Wiseman (2005), Stryszowska (2004), Zeithammer (2005), Peters and Severinov (2006), Rogers, David, Schiff, and Jennings (2007) have sought to explain the bidding behaviour of buyers in online auctions, which include multiple bidding (i.e., successively increasing the amount placed as the maximum bid in the eBay proxy agent, as opposed to bidding the true value) and sniping (i.e., starting to bid near the end of the auction). In fact, as a result of the sniping behaviour of the agents, the English auctions run by eBay can be approximated as Vickrey auctions since bidders no longer have as much information about the status of bidding in the auction. All of these papers, however, focus only on the sequential auction problem, and do not discuss the bidding strategies when multiple auctions close simultaneously. On the other hand, Boutilier, Goldszmidt, and Sabata (1999), Gopal, Thompson, Tung, and Whinston (2005), Juda and Parkes (2006) have explored means to improve the efficiency of simultaneous auctions by either proposing sequential auction mechanisms or the use of options.[4] Whereas these works are aimed at changing or augmenting the auction mechanism to improve their outcome, in our case we take the auctions as given and focus on the strategic aspects from a bidder's perspective.

## 3. Bidding in Multiple Vickrey Auctions

The model consists of $m$ sellers, each of whom acts as an auctioneer. Each seller auctions one item; these items are complete substitutes (i.e., they are equal in terms of value and a bidder obtains no additional benefit from winning more than one of them). The $m$ auctions are executed simultaneously; that is, they end simultaneously and no information about the outcome of any of the auctions becomes available until the bids are placed.[5] However, in Section 5.3 we briefly investigate markets with both sequential and simultaneous auctions since such markets are common in practice, especially online.

Furthermore, we generally assume that all the auctions are identical (i.e., a bidder is indifferent between them), but we relax this assumption in Section 5.2 where the auctions have different valuation distributions and/or numbers of participating bidders and the global bidder may thus prefer one auction over another. Finally, we assume free disposal and that bidders maximise their expected profit. These two assumptions, together with the assumption about the complete substitutes are implicit throughout the paper.

---

4. The notion is similar to financial options, in that these "auction options" provide the buyer the right, but not the obligation, to buy a specific item at a specified price (the strike price) during a specified period of time. However, the analysis of these options differs from the financial perspective due to very different assumptions when modelling them.

5. We note that, although this paper focuses on sealed-bid auctions, where this is the case, the conditions are similar for last-minute bidding or *sniping* in iterative auctions such as eBay (Roth & Ockenfels, 2002); when some of the auctions close at almost the same time, due to network delays there may be insufficient time to obtain the results of one auction before proceeding to bid in the next one.

## 3.1 The Auctions

The seller's auction is implemented as a second-price sealed bid auction, where the highest bidder wins but pays the second-highest price. This format has several advantages for agent-based settings. Firstly, it is communication efficient in terms of the number of interactions, since it requires each bidder to place a single bid once, and the auctioneer to respond once to each bidder with the outcome. In contrast, an iterative auction such as the English auction usually requires several interactions and is typically more time consuming. Secondly, for the single-auction case (i.e., where a bidder places a bid in at most one auction), the optimal strategy is to bid the true value and thus requires no computation (once the valuation of the item is known). This strategy is also weakly dominant (i.e., it is independent of the other bidders' decisions), and therefore it requires no information about the preferences of other agents (such as the distribution of their valuations).

## 3.2 Global and Local Bidders

We distinguish between global and local bidders. The former can bid in any number of auctions, whereas the latter bid in only a single one. Local bidders are assumed to bid according to the weakly dominant strategy and bid their true valuation.[6] We consider two ways of modelling local bidders: *static* and *dynamic*. In the first model, the number of local bidders is assumed to be known and equal to $n$ for each auction. In the latter model, on the other hand, the average number of bidders is equal to $n$, but the exact number is unknown and may vary for each auction. This uncertainty is modelled using a Poisson distribution (more details are provided in Section 4.1).

As we will later show, a global bidder that bids optimally has a higher expected utility compared to a local bidder, even though the items are complete substitutes and a bidder requires only one of them. Nevertheless, we can identify a number of compelling reasons why not all bidders may choose to bid globally:

- **Participation Costs.** Although the bidding itself may be automated by an autonomous agent, it still takes time and/or money, such as entry fees and time to setup an account, to participate in a new auction. In that case, if the marginal benefits from bidding in two auctions instead of a single auctions are less than the participation costs, then a buyer is better off choosing and bidding in only one of the auctions.

- **Information.** Bidders may simply not be aware of other auctions selling the same type of item. Even if this is known, however, a bidder may not have sufficient information about the distribution of the valuations of other bidders and the number of participating bidders. Whereas this information is not required when bidding in a single auction (because of the dominance property in a second-price auction), it is important when bidding in multiple simultaneous auctions. Such information can be obtained by an expert user or be learned over time, but is often not available to a novice.

---

6. Note that, since bidding the true value is optimal for local bidders irrespective of what others are bidding, their strategy is not affected by the presence of global bidders.

- **Risk Attitude.** Although a global bidder obtains a higher utility on average, such a bidder runs a risk of incurring a loss (i.e., a negative utility) when winning multiple auctions. A risk averse bidder may not be willing to take that chance, and so may choose to participate in fewer or even just a single auction to avoid such a potential loss. Whether a bidder chooses to reduce the number of auctions to a single one depends of the degree of risk aversion. In general, we would expect an agent to take less risk and bid in fewer auctions if the stakes are higher, i.e., in the case of high-value transactions. A global bidding agent, on the other hand, could be representing a large firm with sufficient funds to cover any losses, and these agents are more likely to be risk neutral.

- **Budget Constraints.** Related to the previous point, a bidder may not have sufficient funds to take a loss in case it wins more than one auction. In more detail, for a fixed budget $b$, the sum of bids should not exceed $b$, thereby limiting the number of auctions a bidder can participate in and/or lowering the actual bids that are placed in those auctions (see also Section 5.1 where we investigate budget constraints in more detail).

- **Bounded Rationality.** As will become clear from this paper, an optimal strategy for a global bidder is harder to compute than a local one. A bidder will therefore only bid globally if the costs of computing the optimal strategy outweigh the benefits of the additional utility. Moreover, the concept of a local bidder also captures the notion of a 'manual' bidder who does not use an intelligent bidding agent to compute the optimal bid. For a human bidder it is clearly easier to bid their true value in a single auction which requires no calculation at all (again, given that the value is known).

## 4. Identical Simultaneous Auctions

In this section, we provide a theoretical analysis of the optimal bidding strategy when a global bidder participates in identical simultaneous auctions. In particular, we address the case where all other bidders are local and simply bid their true valuation. However, the analysis is more general and applies to any setting where the distribution of the *best competitive bid* is known and satisfies certain properties. After we describe the global bidder's expected utility in Section 4.1, we show in Section 4.2 that it is always optimal for a global bidder to participate in the maximum number of simultaneous auctions available. Subsequently, in Section 4.3 we significantly reduce the complexity of finding the optimal bids for the multi-auction problem in case all auctions are equivalent, and we then apply these methods to find optimal strategies for specific examples.

### 4.1 The Global Bidder's Expected Utility

In what follows, the number of sellers (auctions) is $m \geq 2$ and $M = \{1, \ldots, m\}$ denotes the set of available auctions. Let $G$ denote the cumulative distribution function of the *best competitive bid* in a particular auction and $g$ the corresponding density function. Equiva-lently, $G(b)$ is the probability of winning a specific auction conditional on placing bid $b$ in this auction. We introduce the following assumptions regarding function $G$:

**Assumption 1.** *The cumulative distribution $G(x)$ has bounded support $[0, v_{max}]$, is continuous within this range, and is strictly increasing for $0 < x < v_{max}$.*

A global bidder has a valuation $v \in (0, v_{max}]$ (NB. we do not consider the trivial case where $v = 0$, and we assume that $v$ is within the bounds of $G$) and places a *global bid* **b**, which is a vector that specifies a (possibly different) bid $b_i \in [0, v_{max}]$ for each auction $i \in M$. Now, given the setting described in Section 3 with identical auctions, the expected utility $U$ for a global bidder with global bid **b** and valuation $v$ is given by:

$$U(\mathbf{b}, v) = v \left[ 1 - \prod_{i \in M} (1 - G(b_i)) \right] - \sum_{i \in M} \int_0^{b_i} y g(y) dy. \tag{1}$$

Here, the left part of the equation is the valuation multiplied by the probability that the global bidder wins *at least* one of the $m$ auctions and thus corresponds to the expected benefit. In more detail, note that $(1 - G(b_i))$ is the probability of *not* winning auction $i$ when bidding $b_i$, $\prod_{i \in M}(1 - G(b_i))$ is the probability of not winning any auction, and thus $[1 - \prod_{i \in M}(1 - G(b_i))]$ is the probability of winning at least one auction. The right part of Equation (1) corresponds to the total expected costs or payments. To see the latter, note that the expected payment of a single second-price auction when bidding $b$ equals $\int_0^b y g(y) dy$ and is independent of the expected payments for other auctions.

Now, Equation (1) can be used to address the setting where all bidders except the global bidder are local. This is done as follows. Let the number of *local* bidders be $n \geq 1$. A local bidder's valuation $v \in [0, v_{max}]$ is independently drawn from a cumulative distribution $F$ with probability density $f$, where $F(x)$ has the same properties as $G(x)$ (i.e., has support $[0, v_{max}]$, is continuous within this range, and is strictly positive for $0 < x < v_{max}$). Note that, since it is a dominant strategy for a local bidder to bid the true value, no additional assumptions are needed about the knowledge of these local bidders regarding the distributions of other bidders . In case the local bidders are *static*, i.e., there are exactly $n$ local bidders with equal distributions, $G$ is simply the highest-order statistic $G(b) = F(b)^n$, and $g(b) = dG(b)/db = nF(b)^{n-1}f(b)$. However, we can use the same equation to model *dynamic* local bidders in the following way:

**Lemma 1.** *By replacing the highest-order statistic $G(y)$ and the corresponding density function $g(y)$ with:*

$$\hat{G}(y) = e^{n(F(y)-1)},$$

$$\hat{g}(y) = d\hat{G}(y)/dy = n f(y) e^{n(F(y)-1)},$$

*Equation (1) becomes the expected utility where the number of local bidders in each auction is described by a Poisson distribution with average $n$ (i.e., where the probability that $n$ local bidders participate is given by $P(n) = n^n e^{-n}/n!$).*

PROOF To prove this, we first show that $G(\cdot)$ and $F(\cdot)$ can be modified such that the number of bidders per auction is given by a *binomial* distribution (where a bidder's decision to participate is given by a Bernoulli trial) as follows:

$$G'(y) = F'(y)^N = (1 - p + p\,F(y))^N, \qquad (2)$$

where $p$ is the probability that a bidder participates in the auction, and $N$ is the total number of bidders. To see this, note that not participating is equivalent to bidding zero. As a result, $F'(0) = 1 - p$ since there is a $1 - p$ probability that a bidder bids zero at a specific auction, and $F'(y) = F'(0) + p\,F(y)$ since there is a probability $p$ that a bidder bids according to the original distribution $F(y)$. Now, the average number of participating bidders is given by $n = p\,N$. By replacing $p$ with $n/N$, Equation (2) becomes $G'(y) = (1 - n/N + (n/N)F(y))^N$. Note that a Poisson distribution is given by the limit of a binomial distribution. By keeping $n$ constant and taking the limit $N \to \infty$, we then obtain $G'(y) = e^{n(F(y)-1)} = \hat{G}(y)$. $\qquad\square$

The results that follow apply to both the static and dynamic model unless stated otherwise.

## 4.2 Participation in Multiple Auctions

We now show that, for any valuation $0 < v < v_{max}$, a utility-maximising global bidder should always place non-zero bids in all available auctions.[7] To prove this, we show that the expected utility increases when placing an arbitrarily small bid compared to not participating in an auction. This holds even when the auctions are not identical. In the following let $\mathbf{b}_{b_j=\hat{b}_j}$ denote the global bid $\mathbf{b}$ where the $j^{th}$ bid $b_j = \hat{b}_j$. More formally,

**Theorem 1.** *Under Assumption 1 and given that the global bidder has a valuation $0 < v < v_{max}$, consider a global bid $\mathbf{b}$ with $b_i \le v$ for all $i \in M$. Suppose that $b_j = 0$ for some auction $j \in M$, then Equation (1) is not maximised, i.e., there exists a $\hat{b}_j > 0$ such that $U(\mathbf{b}_{b_j=\hat{b}_j}, v) > U(\mathbf{b}, v)$.*

PROOF We need to show that there exists a $\hat{b}_j > 0$ such that $U(\mathbf{b}_{b_j=\hat{b}_j}, v) - U(\mathbf{b}, v) > 0$. Using Equation (1), the marginal expected utility for participating in auction $j$ w.r.t. bidding zero in that auction can be written as:

$$U(\mathbf{b}_{b_j=\hat{b}_j}, v) - U(\mathbf{b}, v) = vG(\hat{b}_j) \prod_{i \in M \setminus \{j\}} (1 - G(b_i)) - \int_0^{\hat{b}_j} yg(y)dy.$$

Now, using integration by parts, we have $\int_0^{\hat{b}_j} yg(y) = \hat{b}_j G(\hat{b}_j) - \int_0^{\hat{b}_j} G(y)dy$ and the above equation can be rewritten as:

$$U(\mathbf{b}_{b_j=\hat{b}_j}, v) - U(\mathbf{b}, v) = G(\hat{b}_j)\left[ v \prod_{i \in M \setminus \{j\}} (1 - G(b_i)) - \hat{b}_j \right] + \int_0^{\hat{b}_j} G(y)dy. \qquad (3)$$

---

7. We note that this does not necessarily hold in the boundary case where $v = v_{max}$. However, in practice, we find that even here the optimal strategy is to bid below the true value in multiple auctions instead of the true value in a single auction. This is especially the case when the number auctions is large (see also Section 4.3.4).

Clearly, since $\hat{b}_j > 0$ and $g(x) > 0$ for $x > 0$ (due to Assumption 1) we have that $G(\hat{b}_j)$ and $\int_0^{\hat{b}_j} G(y)dy$ are both strictly positive. Moreover, given that $b_i \leq v < v_{max}$ for $i \in M$ and that $v > 0$, it follows that $v \prod_{i \in M \setminus \{j\}} (1 - G(b_i)) > 0$. Now, suppose we set $\hat{b}_j = \frac{1}{2} v \prod_{i \in M \setminus \{j\}} (1 - G(b_i))$, then $U(\mathbf{b}_{b_j = \hat{b}_j}, v) - U(\mathbf{b}, v) = G(\hat{b}_j) \left[ \frac{1}{2} v \prod_{i \in M \setminus \{j\}} (1 - G(b_i)) \right] + \int_0^{\hat{b}_j} G(y)dy > 0$ and thus $SU(\mathbf{b}_{b_j = \hat{b}_j}, v) > U(\mathbf{b}, v)$. □

The above proof applies to the setting where auctions are identical. However, it is easy to see that the same argument holds if $G$ differs for each auction.

This result states that, even though there is risk of winning and having to pay for more than one item (and a buyer disposes of any additional items won), the best strategy is to participate in all auctions. Therefore, in expectation, the increasing probability of winning a single item outweighs the possible loss incurred when winning more than one of them. To obtain a better understanding of why this is true, consider the following more intuitive argument. Suppose that a global bidder bids in $k < m$ auctions. In that case, there is some non-zero probability that the bidder wins none of the auctions, and thus there is a non-zero expected demand for at least one of the items in the remaining auctions. Since this argument holds for any $k < m$, by induction a global bidder should bid in all $m$ auctions.

Note that Theorem 1 holds only when $v$ is strictly smaller than $v_{max}$. In the case that $v = v_{max}$ there are two possibilities: either it is optimal to bid $v_{max}$ in one auction in which case the bids in the other auctions should be zero (since the bidder is guaranteed to win), or it is optimal to bid below $v_{max}$ but strictly positive in all auctions. As we will show in Section 4.3.4 we find empirically that the first is true only when the number of auctions is small, and the latter is the case for large numbers of auctions.

### 4.3 The Optimal Global Bid

A general solution to the optimal global bid requires the maximisation of Equation (1) in $m$ dimensions, an arduous task, even when applying numerical methods. In this section, however, we show how to reduce the entire bid space to a single dimension in most cases, thereby significantly simplifying the problem at hand. First, however, in order to find the optimal solutions to Equation (1), we set the partial derivatives to zero:

$$\frac{\partial U}{\partial b_i} = g(b_i) \left[ v \prod_{j \in M \setminus \{i\}} (1 - G(b_j)) - b_i \right] = 0. \tag{4}$$

Now, equality (4) holds either when $g(b_i) = 0$ or when $\prod_{b_j \in \mathbf{b} \setminus \{b_i\}} (1 - G(b_j))v - b_i = 0$. In the model with dynamic local bidders, $g(b_i)$ is always greater than zero, and can therefore be ignored (since $g(0) = nf(0)e^{-n}$ and we assume $f(y) > 0$). In other cases, $g(b_i) = 0$ only when $b_i = 0$. However, Theorem 1 shows that the optimal bid is non-zero for $0 < v < v_{max}$. Therefore, we can ignore the first part, and the second part yields:

$$b_i = v \prod_{j \in M \setminus \{i\}} (1 - G(b_j)). \tag{5}$$

In other words, the optimal bid in auction $i$ is equal to the bidder's valuation multiplied by the *probability of not winning any of the other auctions*. It is straightforward to show that the second partial derivative is negative, confirming that the solution is indeed a maximum when keeping all other bids constant. Moreover, since the optimal bid requires that $0 < b_i < v_{max}$ for $v < v_{max}$ (due to Theorem 1 and since bidding more than the valuation is clearly suboptimal), we only need to consider interior solutions. Thus, Equation (5) provides a means to derive the optimal bid for auction $i$, given the bids in all other auctions.

Now, by taking the partial derivatives for each auction and rewriting Equation (5), the optimal global bid must obey the following relationship: $b_1(1 - G(b_1)) = b_2(1 - G(b_2)) = \ldots = b_m(1 - G(b_m))$. By defining $H(b) = b(1 - G(b))$ we can rewrite the equation to:

$$H(b_1) = H(b_2) = \ldots = H(b_m) = v \prod_{j \in M} (1 - G(b_j)). \tag{6}$$

In what follows, we apply Equation (6) to reduce the search space, and first show that, for a large class of probability distributions, the optimal global bid consists of at most two different values, thereby reducing the search space to two dimensions. In Section 4.3.2 we further reduce this to a single dimension. In Section 4.3.3 we consider limit results when the number of auctions goes to infinity. Finally, in Section 4.3.4 we perform a numerical analysis of the optimal bidding strategy for specific cases when the number of auctions is finite, and we consider to what extent a global bidder benefits compared to bidding locally.

### 4.3.1 REDUCING THE SEARCH SPACE

In this section, we first show that the optimal global bids consists of at most two different values when the function $H(b) = b(1 - G(b))$ has a unique critical point. More formally, we introduce the following requirement:

**Assumption 2.** $H(b) = b(1 - G(b))$ *has a unique critical point* $b^f$, *i.e., there exists a* $b^f$ *s.t.* $\frac{d}{db}H(b^f) = 0$ *and for all* $b \neq b^f$, $\frac{d}{db}H(b) \neq 0$.

We then go on to show that this requirement is met for a wide class of distributions which are characterised by a *non-decreasing hazard rate*. Let the two bid values be denoted by $b_-$ and $b_+$, where $b_- \leq b_+$. Formally:

**Theorem 2.** *Under Assumptions 1 and 2, the global bid* $\mathbf{b}^*$ *maximising Equation (1) contains at most two distinct bid values:* $b_-$ *and* $b_+$. *In addition,* $b_- \leq b^f \leq b_+$, *where* $b^f$ *is the unique critical point of* $H(b) = b(1 - G(b))$.

PROOF Suppose $H$ has a unique critical point, $b^f$. $H(0) = H(v_{max}) = 0$, and $H(b) > 0$ in $(0, v_{max})$, so $b^f$ must in fact be the global maximum of $H$. Furthermore, for $b < b^f$ $H$ is strictly increasing while for $b > b^f$ $H$ is strictly decreasing. This also implies that $H(x) = y$ has at most two solutions: if $y > H(b^f)$ then there are no solutions, since $b^f$ is the global maximiser of $H$; if $y = H(b^f)$ there is a unique solution, namely $b^f$; if $y < H(b^f)$ then applying the intermediate value theorem to $H$ on the interval $[0, b^f]$ gives one solution, namely $b_- \leq b^f$, and on the interval $[b^f, v_{max}]$ gives the other, namely $b_+ \geq b^f$. There can be only one solution in each interval because $H$ is strictly monotonic on each.

Now, Equation (6) implies that $H(b_i)$ is equal for all $i \in M$. Therefore, given that $H(x) = y$ has at most two solutions, there can be at most two distinct interior bids $b_i$, namely $b_-$ and $b_+$. As mentioned before, from Theorem 1 the utility maximizing solution has $0 < b_i < v_{max}$ when $v < v_{max}$, and therefore all solutions are interior ones. In the case that $v = v_{max}$ either $b_+ = v_{max}$ and $b_- = 0$, or we have as before that $b_+ < v_{max}$ and $b_- > 0$. $\qquad\square$

We now show that the unique critical point of $H$ is guaranteed if $G$ has a non-increasing hazard rate within the interval $[0, v_{max}]$. We choose this property since it encompasses a large number of distributions, including those with log-concave density functions such as uniform, normal and exponential (we refer to the work by Barlow, Marshall, and Proschan, 1993 and Bergstrom and Bagnoli, 2005 for a list of such functions). Formally, a hazard rate (see e.g., Krishna, 2002) of a cumulative distribution function $F$ is denoted by $\lambda_F$ and is defined by:

$$\lambda_F(x) \equiv \frac{f(x)}{1 - F(x)}.$$

We have the following result:

**Lemma 2.** *Under Assumption 1, if $\lambda_G(b)$ is non-decreasing in $(0, v_{max})$, then the function $H(b) = b(1 - G(b))$ has a unique critical point $b^f$ in that interval.*

PROOF See Appendix A.1.

We now extend the proof to the distribution functions of the local bidders $F$ by showing that, if the hazard rate of $F$ is non-decreasing, the hazard rate of $G$ is also non-decreasing, and thus the reduction also applies. This holds both for local and static bidders. Formally:

**Theorem 3.** *If $\lambda_F(b)$ is non-decreasing, then $\lambda_G(b)$ and $\lambda_{\widehat{G}}(b)$ are also non-decreasing, where $G(b) = F(b)^n$ and $\widehat{G} = e^{n(F(y)-1)}$ for $n \geq 2$.*

PROOF See Appendix A.1.

### 4.3.2 Characterising the Optimal Bids

Using the above results we are able to reduce the optimal global bid to two values, a *high* bid, $b_+ \geq b^f$, and a *low* bid, $b_- \leq b^f$. However, we do not know the number of auctions in which to bid high and low. In this section we show that it is optimal to bid high in at most one auction, and low in all other auctions. Using this result, we can then write the high bid in terms of the low bids, reducing the search space even further.

**Theorem 4.** *Under Assumptions 1 and 2, the global bid $\mathbf{b}^*$ that maximises Equation 1 has at most one high bid, i.e., at most one $i \in M$ for which $b_i > b^f$, where $b^f$ is the unique critical point of $H(b) = b(1 - G(b))$.*

PROOF See Appendix A.1.

Together with Lemma 2 this implies that, for distributions with non-decreasing hazard rates, either there exists one high bid and $m - 1$ low bids, or all bids are low. Note that we can now calculate the value of the high bid analytically given the low-bid value by using Equation (5). Consequently, finding the optimal global bid reduces to optimising a single variable (i.e., the value for the low bids). This value can then be computed numerically using standard optimisation techniques such as the Quasi-Newton method, or, alternatively, the bids can be discretised and a brute-force search can be applied to find the optimum. Whatever method is selected it is important to notice that, due to the reduction of the search space, the *computational complexity* of calculating the optimal outcome numerically is independent of the number of auctions (or indeed the number of bidders).

The result of Theorem 4 suggests that it is optimal to restrict attention to a single auction by bidding high in that auction, and then using the remaining auctions as a backup in case the high-bid auction fails. As we will show in Section 4.3.4 this is often the case in practise when the number of auctions is small. When the number of auctions is large, however, we find that it is optimal to bid low in all of the auctions, irrespective of the bidder's valuation. We derive theoretical results for the limit case where the number of auctions goes to infinity in Section 4.3.3, and then consider empirical results for the finite case in Section 4.3.4.

### 4.3.3 Limit Results

In this section we investigate how the optimal bidding changes as the number of auctions becomes very large and consider whether there are general patterns that characterise the optimal strategy. The first, most basic result is that, as the number of auctions increases, the agent is able to extract an increasingly greater utility and this approaches the maximum possible utility, $v$. To prove this, without loss of generality we restrict the strategy of the global bidder by only considering uniform bidding (i.e., where all bids are equal). Let $\mathbf{b}_m^{u*}$ denote the optimal global bid when bidding in $m$ auctions when the bidder is confined to using uniform bids:

**Theorem 5.** *Under Assumption 1, the expected utility as defined by Equation 1 from playing the optimal uniform bid $\mathbf{b}_m^{u*}$ converges to $v$, in the sense that for all $\epsilon > 0$ there is a constant $m_\epsilon$ such that $m > m_\epsilon$ implies $U(\mathbf{b}_m^{u*}) > v - \epsilon$.*

PROOF See Appendix A.2.

Note that, since $v$ is an upper bound on the utility that can be achieved by any global bidder, this result implies that, as the number of auctions increases, eventually uniform bidding will always be superior to non-uniform bidding. More formally:

**Corollary 1.** *Under Assumption 1, for sufficiently large $m$ the globally optimal bid $\mathbf{b}^*$ maximising Equation (1) is equal to the optimal uniform bid $\mathbf{b}_m^{u*}$, independent of $v$.*

PROOF This corollary follows from Theorem 5 with $\epsilon = EP(b^f) = \int_0^{b^f} xg(x)\,dx$: for $m > m_\epsilon$ uniform bidding gives utility at least $v - \epsilon$, whereas non-uniform bidding gives strictly less:

$$U(\mathbf{b}^*) < v - EP(b_+) < v - EP(b^f) = v - \epsilon.$$

$\square$

In practice, we find that it is not necessary for the number of auctions to be particularly large before uniform bidding is optimal for all $v$. To this end, in the next section we provide examples of optimal bidding strategy for specific settings.

### 4.3.4 EMPIRICAL EVALUATION

In this section, we present results from an empirical study and characterise the optimal global bid for specific cases with finite numbers of auctions. Furthermore, we measure the actual utility improvement that can be obtained when using the global strategy. The results presented here are based on a uniform distribution of the valuations with $v_{max} = 1$, and the static local bidder model, but they generalise to the dynamic model and other distributions. Figure 1 illustrates the optimal global bids and the corresponding expected utility for various $m$ and $n = 5$, but again the bid curves for different values of $m$ and $n$ follow a very similar pattern.

As shown in Figure 1, for bidders with a relatively low valuation, the optimal strategy is to submit $m$ equal bids at, or very close to, the true value. After the valuation reaches a certain point, however, placing equal bids is no longer the optimal strategy when the number of auctions is small (in Figure 1 this occurs at $m = 4$ and $m = 6$). At this point, a so-called pitchfork bifurcation is observed and the optimal bids split into two values: a single high bid and $m - 1$ low ones. In all experiments, however, we consistently observe that the optimal strategy is always to place uniform bids when the valuation is relatively low. Moreover, the bifurcation point moves to the right as $m$ increases, and disappears altogether when $m$ becomes sufficiently large ($m \geq 10$ in Figure 1) at which point the optimal bids are uniform (note that this holds even when $v = v_{max}$). Note also that the uniform optimal bids move closer to zero as $m$ tends to infinity.

As illustrated in Figure 1, the utility of a global bidder becomes progressively higher with more auctions. Note that, consistent with the limit results from Section 4.3.3, the utility approaches the upper bound $v$ as the number auctions becomes large. In absolute terms, the improvement is especially high for bidders that have an above average valuation, but not too close to $v_{max}$. The bidders in this range thus benefit most from bidding globally. This is because bidders with very low valuations have a very small chance of winning any auction, whereas bidders with a very high valuation have a high probability of winning a single auction and benefit less from participating in more auctions. In contrast, as shown in Figure 1, if we consider the utility *relative* to bidding in a single auction, this is much higher for bidders with relatively low valuations. In particular, we notice that a global bidder with a low valuation can improve its utility by up to $m$ times the expected utility of bidding locally. Intuitively, this is because the chance of winning one of the auctions increases by up to a factor $m$, whereas the increase in the expected cost is negligible. For high valuation buyers, however, the benefit is not that obvious because the chances of winning are relatively high even in case of a single auction.

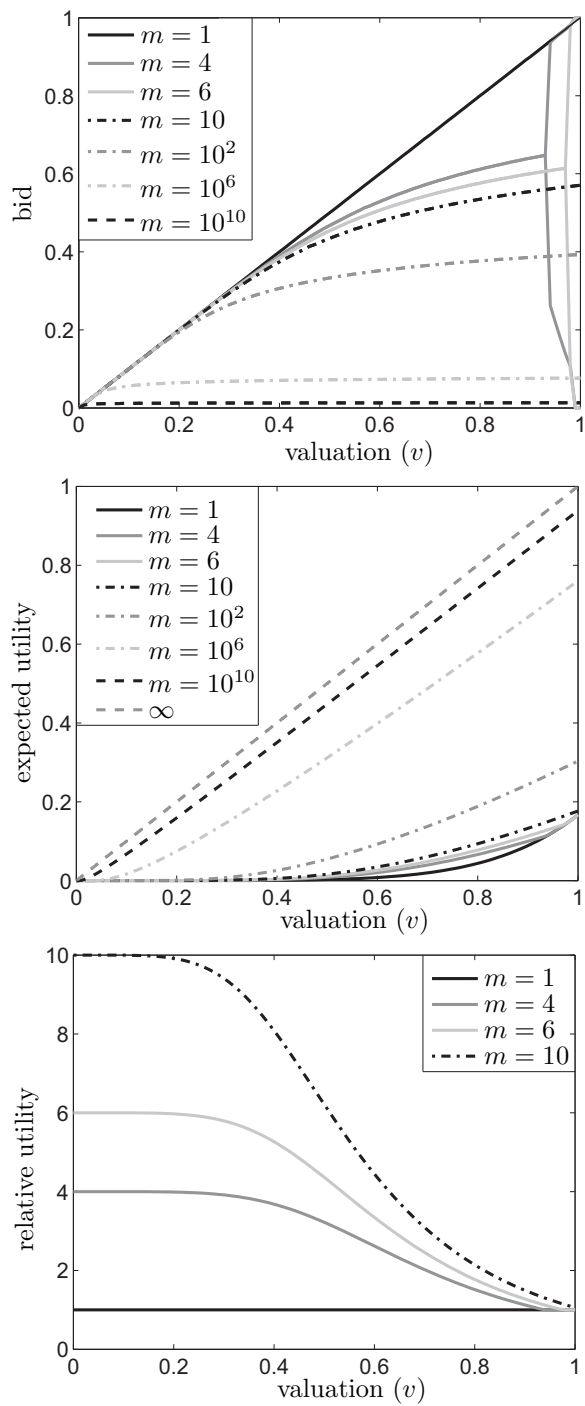

Figure 1: The optimal global bid, the corresponding expected utility, and the expected utility proportional to that of a local bidder for a setting with $n = 5$ static local bidders and varying number of auctions ($m$). Note that the results for $m = 1$ correspond to those of a local bidder. For comparison, the expected utility is also shown as the number of auctions approaches infinity.

## 5. Extensions: Budgets, Non-Identical and Sequential Auctions

In the previous section, we considered the best response strategy when a global bidder faces multiple identical simultaneous auctions and has no financial constraints. In this section, we generalise the results to other settings. In particular, we investigate three important extensions:

1. In Section 5.1 we investigate a setting where the global bidder has a limited budget which constrains the sum of bids or *exposure*.

2. In Section 5.2 we investigate a setting where the auctions differ in their probability of winning. Such differences arise, for example, when the auctions have different numbers of local bidders participating.

3. In Section 5.3 we extend the results to sequential auctions and consider a setting where the resources of interest are auctioned both simultaneously and sequentially.

### 5.1 Budget-Constrained Bidding

The derivation of the optimal global bid has shown that it is an optimal strategy for the global bidder to bid in all the simultaneous auctions. Now, such a strategy implicitly assumes that the bidder does not face financial constraints (i.e., a bidder can pay for all the items won). However, often a bidder has limited resources which may restrict its bidding strategy. In this section we study how a budget can limit the space of possible strategies available to a global bidder and affect its optimal strategy. In particular, we consider the case where a budget constrains the *exposure*, i.e., the sum of the bids.[8] This occurs, for example, when the global bidder has limited liquidity and faces very negative consequences when it cannot pay for all the items it wins (e.g., going bankrupt or being thrown out of the system).

Now, the importance of taking budget constraints into account becomes even more pronounced from the following result which shows that, as the number of auctions increases, the required budget or exposure in the unconstrained case will exceed any given limit:[9]

**Theorem 6.** *Under Assumption 1 and assuming that $g(b)$ is bounded throughout $[0, v_{max}]$, for all $v > 0$ and for any $C$, there exists an $m$ for which the exposure of the optimal global bid $\mathbf{b}^*$ maximising Equation (1) exceeds $C$, i.e., $\sum_{i \in M} b_i > C$.*

PROOF See Appendix A.3.

Thus, despite the low probability of the bidder having to pay the sum of its bids (especially, given that each auction is a second-price auction), in practice a bidder may still want to limit this amount.

---

8. We note that, although in practice a budget constraint is on the payments rather than the bids, even in the second-price auctions the worst-case outcome is that a bidder pays their bids in all auctions, and therefore a hard budget constraint is equivalent to constraining the sum of bids.

9. Note that this occurs despite the fact that bids tend to zero as m goes to infinity. However, the exposure does not (see the proof of Theorem 5).

In more detail, we can formulate the budget-constrained problem faced by the bidder as:

$$\max_{\mathbf{b} \in [0,v]^m} U(\mathbf{b}, v) \qquad s.t. \sum_{i \in M} b_i \leq C, \tag{7}$$

where $C$ is the budget limit and $U(\mathbf{b}, v)$ is the utility of the global bidder as given by Equation (1). We consider budget constraints by distinguishing between three cases. In the following, $\mathbf{b}^* = (b_1^*, \ldots, b_m^*)$ refers to the unconstrained optimal solution (i.e., the optimal solution in the absence of any budget constraints), whereas $\mathbf{b}^{c*} = (b_1^{c*}, \ldots, b_m^{c*})$ refers to the optimal global bid subject to budget constraints.

**Case (1)** $\sum_{i \in M} b_i^* > C$ and $v \geq C$.

Here, the sum of the unconstrained optimal bids exceeds the budget and it is therefore required to recompute the optimal bid given the constraint. Moreover, the budget constraint is equal or less than the valuation. For this case, we are able to show that it is a best strategy to bid in a *single* auction and to place bid $b_i = C$ in this auction for fairly general probability density functions of $g$. By so doing, this provides one of the justifications for the existence of *local bidders* (as outlined in Section 3.2). Although this result is intuitive, it is not straightforward since a bidder may still decide to divide its budget across several auctions. Below we provide the proof of this result.

**Case (2)** $\sum_{i \in M} b_i^* > C$ and $v < C$.

As in the previous case, we need to recompute the optimal bid. However, as we will show, the optimal strategy for the global bidder in this case is to bid in multiple auctions (as in the unconstrained case). Moreover, in contrast to the unconstrained case, the global bid may consist of more than two different values.

**Case (3)** $\sum_{i \in M} b_i^* \leq C$.

In this case, the sum of the unconstrained optimal bids is less than the budget. Thus the result is trivial and we have $\mathbf{b}^* = \mathbf{b}^{c*}$.

Examples of these three different cases are depicted in Figure 2. This figure shows the optimal strategy for bidding in 4 simultaneous auctions when local bidders have uniformly distributed valuations within the range $[0, 1]$ and the global bidder has a budget constraint $C = 0.8$.[10] Clearly, for a global bidder with a low valuation the budget constraint does not affect the optimal bidding strategy (case 3). Case 1 occurs when the bidder has a valuation at or above 0.8, and case 2 occurs in the in-between range. As this figure shows, the optimal strategy in the latter two cases is qualitatively very different from the unconstrained case; whereas the unconstrained optimal strategy is to bid high in at most a single auction, now this consists of at most a single *low* bid, several high bids, and placing zero in the remaining auctions. As the budget becomes tighter relative to the valuation, the number of auctions in which the bidder participates decreases until a single auction remains (see Figure 2, case 1). In what follows, we first consider conditions under which such behaviour is optimal for case 1, and subsequently we address case 2 in more detail. Case 3 is trivial and is therefore not considered further.

---

10. Similar patterns are observed in the optimal strategy with varying number of auctions and budget. These patterns can always be grouped in these three cases.

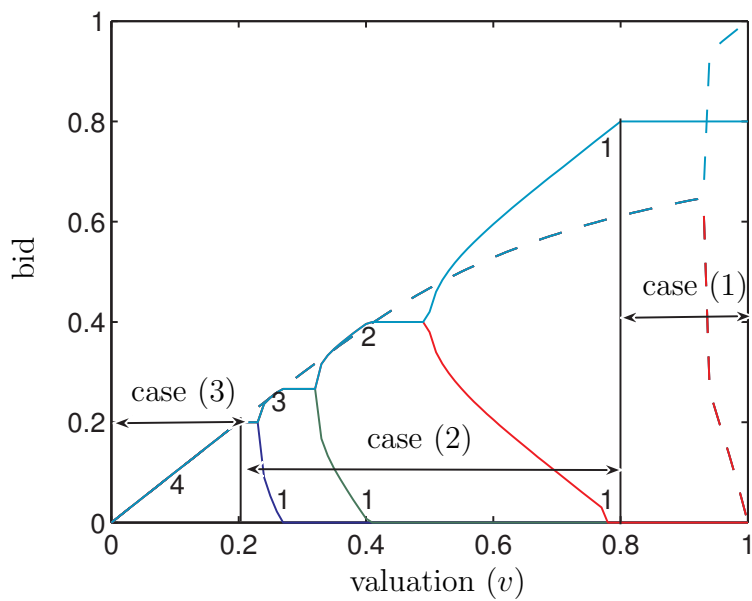

Figure 2: The solid lines denote the optimal bids of a global bidder with budget $C = 0.8$ in a setting where n=5 and m=4. Here, the numbers indicate the *number of auctions* in which a certain bid is placed. As the valuation increases (and the budget remains constant), the bids in the auctions taper off one by one, until a single high bid remains and all other bids are zero. The dotted line represents the unconstrained solution.

We now provide a formal result for the optimal strategy in case 1 and show that, when the budget constraint imposed on a global bidder is equal to or less than the value it attaches to the item it wishes to acquire, it is a best strategy to act as a local bidder under the following conditions:

**Assumption 3.** *The probability density function of the best competitive bid $g(x)$ is convex and $g(0) = 0$.*

The result is stated formally as follows:

**Theorem 7.** *Under Assumptions 1 and 3, if the global bidder has a budget $C \leq v$, then Equation (7) is satisfied for $\mathbf{b}^{c*} = (C, 0, \ldots, 0)$, i.e., it is optimal to bid $C$ in exactly one auction.*

PROOF See Appendix A.3.

Intuitively, the convexity of $g(x)$ is necessary to ensure that the probability of winning an auction increases sufficiently quickly as the bid increases. This way, a higher bid in a single auction results in a higher probability of winning compared to dividing the same amount over several auctions. Although placing a higher bid in a single auction may lead to a higher expected payment, the proof shows that the utility obtained from an increased probability of winning outweighs the expected payment increase. At the same time, the condition $g(0) = 0$ is important, otherwise the probability of winning increases sufficiently quickly such that it is optimal to spread the budget over multiple auctions. The importance of the two conditions becomes apparent by the following corollary which shows that, for the special case where $C = v$, the condition that $g(0) = 0$ is in fact a *necessary* condition for bidding in a single auction to be optimal, and, furthermore, that this strategy is no longer optimal in case $g(x)$ is concave.

**Corollary 2.** *Under Assumption 1, if either $g(0) > 0$ or $g(x)$ is strictly concave, and the global bidder has a budget $C = v$, then Equation (7) is satisfied by bidding strictly positive in at least two auctions.*

Proof See Appendix A.3.

Note that, in case of static local bidders, the condition that $g(x)$ is convex holds, for example, when $f(x)$ is convex, increasing, and non-negative (and thus $F(x)$ is also convex), since $g(x) = nF^{n-1}(x)f(x)$. However, the condition holds in other more general cases as well; especially if the number of local bidders is high, a concave local bidder distribution can easily result in a convex $g(x)$. Also note that the condition $g(0) = 0$ holds in the case of static local bidders and $n \geq 2$, even if $f(0) > 0$ (but not in the case of dynamic local bidders unless $f(0) = 0$). Finally, we note that convexity of $g(x)$ implies a non-decreasing hazard rate and therefore the conditions are stronger than those imposed by Assumption 2 in Section 4.3.

We now move on to consider cases 2 and 3. Case 2 cannot be analysed as easily as case 1. However, a good insight into the constrained optimal strategy can be obtained by considering the Lagrangian of Equation (7). Since the budget constraint is an *in*equality constraint, in addition to the regular Langrangian multiplier we need to introduce the slack variable $\delta$ to convert the inequality into an equality. The variable is first squared to ensure a positive value and is then added to the constraint which becomes $C - \sum_{i \in M} b_i + \delta^2 = 0$. The Langrange function becomes as follows:

$$\phi(B, \lambda, \delta) = U(\mathbf{b}, v) + \lambda \left( \sum_{i \in M} b_i - C + \delta^2 \right). \tag{8}$$

By setting partial derivatives to zero, this results in the following $m + 2$ equations to be solved:

$$\begin{aligned}
\frac{\partial \phi(.)}{\partial \lambda} &= \sum_{i \in M} b_i - C + \delta^2 = 0, \\
\frac{\partial \phi(.)}{\partial \delta} &= 2\lambda\delta = 0, \\
\frac{\partial \phi(.)}{\partial b_i} &= \frac{\partial U(\mathbf{b}, v)}{\partial b_i} + \lambda = 0 \quad \forall i \in M.
\end{aligned} \tag{9}$$

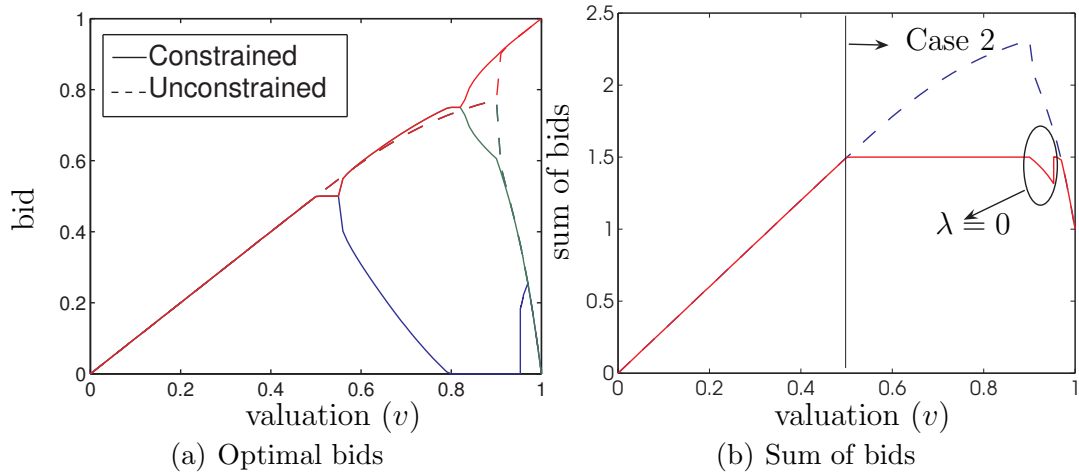

(a) Optimal bids                (b) Sum of bids

Figure 3: The optimal global bid and the corresponding sum of bids both for the unconstrained case and in the case of a budget constraint $C = 1.5$. Here, n=10 and m=3.

From these equations, it can readily be observed that for case 1, $\delta = 0$ thereby leading to the case where the solution of Equation (9) forces the total budget to be spent, i.e., $\sum_{i \in M} b_i = C$.

For case 2, either $\lambda = 0$ corresponding to a local maximum of Equation (4), in which the total budget is not spent or $\delta = 0$ whereby the total budget is spent. These two possible situations are highlighted in Figure 3, showing the optimal global bidding strategy for $n = 10$, $m = 3$ and budget $C = 1.5$, and the corresponding sum of bids (note that here case 1 does not arise since the budget exceeds the highest valuation). The unconstrained solution is also provided for comparison. This example shows that the total amount spent is not necessarily equal to the available budget, even when the unconstrained optimal solution exceeds the budget. Thus, for case 2 we cannot solely consider solutions whereby the sum of bids is equal to the budget since it may be the case that bidding less than the budget yields a greater utility (i.e., when $\lambda = 0$).

Furthermore, as can be observed from Figures 2 and 3(a), the introduction of a budget constraint changes the shape of the optimal strategy for cases 1 and 2. Recall from Section 4.3 that the optimal strategy in the unconstrained case is to either bid equally in all auctions or to bid high in one auction and low and equally in all of the remaining ones. However, with a budget constraint, we now have a region where the best strategy is to bid high in more than one auction and low in the remaining ones. Moreover, parts of the solution now consist of three different bids: high, low, and zero. Hence, Equation (5) no longer holds in case of budget constraints, and the structure of bids observed in Section 4.3 need not be satisfied. This means that we cannot apply the same reduction in search space to efficiently compute the optimal strategy in the case of budget constraints for case 2, and the complexity of computing the optimum using brute force increases exponentially with the number of auctions.

## 5.2 Non-Identical Auctions

Whereas previously we assumed all auctions to be equivalent, we now relax this assumption to the more general case with non-identical auctions. Here, we assume that these auctions differ not in the global bidder's valuation for the item that is sold, but rather in the probability of obtaining the item at a given bid. These differences arise, for example, when the number of bidders and/or the local bidders' valuation distribution vary from one auction to another. This can be used in practice when specific information about the individual auctions is available.[11] In more detail, we assume that each auction $i \in M$ has an individual cumulative distribution function, denoted by $G_i(b)$, and a corresponding density function $g_i(b)$. Now, the expected utility is given by:

$$U(\mathbf{b}, v) = v \left[ 1 - \prod_{i \in M} (1 - G_i(b_i)) \right] - \sum_{i \in M} \int_0^{b_i} y g_i(y) dy, \tag{10}$$

where $\mathbf{b} = (b_1, \ldots, b_m)$ is the global bid, specifying a bid for each auction as before.

It is easy to see that Theorem 1 in Section 4.2 extends to the setting with non-identical auctions and that the optimal strategy is to bid a strictly positive amount in each auction, provided $G_i(b)$ is strictly positive for all $i \in M, 0 < b \leq v_{max}$. But, the auctions are now non-identical and this complicates the bidding space. We revisit the search space problem for non-identical auctions in Section 5.2.1 where a different method is employed to reduce the bidding space. In Section 5.2.2, we show that a global bidder should always bid higher in a more preferred auction. The optimal bidding strategy for a number of non-identical auctions is empirically evaluated in Section 5.2.3.

### 5.2.1 CALCULATING THE OPTIMAL GLOBAL BID

Although a reduction of the search space as described in Section 4.3 is no longer possible when the auctions are not identical, we can still significantly reduce the computation needed to find the optimal bid. To this end, we first set the partial derivatives $\partial U/\partial b_i$ to zero for $i \in M$. Given that $g_i(b_i) > 0$ we have:

$$b_i = v \prod_{k \in M \setminus \{i\}} (1 - G_k(b_k)). \tag{11}$$

By combining the partial derivatives, we then obtain the following relationship:

$$b_1 (1 - G_1(b_1)) = b_2 (1 - G_2(b_2)) = \ldots = b_m (1 - G_m(b_m)) = v \prod_{k \in M} (1 - G_k(b_k)). \tag{12}$$

Generally, it is not possible to find the optimal global bid analytically. Using expression (12), however, we can confine our search to only one of the bids and then, for *each value* of this bid, determine the values of $m - 2$ other bids by the computationally less demanding one-dimensional root-finding operations (see e.g., Burden & Faires, 2004). The remaining bid value can be found analytically using Equation (11).

---

11. For example, eBay auctions reveal the number of visits to web pages of particular items, which can be used to estimate the number of participating bidders at individual auctions (see `http://www.ebay.com` for examples).

To clarify this reduction in search space, we show how this can be applied to a discrete bid space and using a brute-force search approach. To find the optimal (discrete) global bid, we iterate through the discrete space of one of the bids, say $b_1$. For each value of $b_1$ we then find the corresponding $m-1$ bid values such that expression (12) is satisfied as follows. We first calculate $b_1\left(1-G_1(b_1)\right)$ and then search through all discrete values of $b_2$ such that $b_1\left(1-G_1(b_1)\right) = b_2\left(1-G_2(b_2)\right)$.[12] Typically there will be at least two values of $b_2$ for which this equality holds.[13] All the solutions are stored into memory and this is repeated for each $b_i < b_m$. The value for $b_m$ is calculated using Equation (11). This way we can calculate the expected utility *given* $b_1$ such that expression (12) is satisfied. The optimal solution is then found by maximising the expected utility across all possible values of $b_1$ and all combinations of solutions which have been stored into memory. Note that the amount of computation required to find the expected utility for a single value of $b_1$ increases linearly with the number of auctions. However, the number of combinations increases exponentially with the number of auctions. Nevertheless, since the base of this exponential is typically 2 (in particular in the case of non-decreasing hazard rates, see Footnote 13), computation remains tractable when the number of auctions is relatively small but becomes intractable for very large settings.[14] In specific cases, however, where one auction is clearly 'better' than another auction in a precise sense, it is possible to reduce the number of combinations and thus the required computation becomes linear with the number of auctions. This issue is addressed in the next section.

### 5.2.2 Preferred Auctions and their Optimal Bids

In many cases, it is possible to find auctions which are more favourable than others in terms of their expected utility. For example, all else being equal, a bidder is expected to do better on average in auctions with fewer other bidders. A bidder would therefore prefer such auctions over less profitable ones, irrespective of a bidder's own valuation. In this section, we formalise the notion of a preferred auction, and investigate the optimal bids when bidding in multiple auctions with respect to these preferences over auctions.

We use the concept of *stochastic orders* (Shaked & Shanthikumar, 1994) to rank the auctions in terms of a global bidder's preferences. Formally, auction $j$ stochastically dominates auction $i$ when $G_j(b) \leq G_i(b)$ for all $0 \leq b \leq v_{max}$ (Krishna, 2002, Appendix B). The expected utility is then at least as high when bidding *only* in auction $i$ compared to bidding the same amount in auction $j$. This is shown as follows. We can write the expected utility $U$ of auction $i/j$ as: $U_{i/j} = (v - b)G_{i/j}(b) + \int_0^b G_{i/j}(x)dx$. Since both $G_i(b) \geq G_j(b)$ and $\int_0^b G_i(x)dx \geq \int_0^b G_j(x)dx$ when auction $j$ dominates auction $i$, the expected utility of bidding in auction $i$ is at least as high as in auction $j$. We therefore refer to auction $i$ as the (weakly) preferred auction. Furthermore, we say that auction $i$ is *strictly* preferred over auction $j$ at some $b$ if $G_i(b) > G_j(b)$.

---

12. In case of discrete bids this equality rarely holds exactly but this can be resolved by minimising the difference instead, i.e., minimising $b_1\left(1-G_1(b_1)\right) - b_2\left(1-G_2(b_2)\right)$.

13. More precisely, if the function $b\left(1-G(b)\right)$ has a single critical value, then there will be at most two solutions. As shown earlier by Lemma 2 this is the case when $G$ has a non-decreasing hazard rate. Note that, in case of discrete bids, these solutions are local minima when taking the difference.

14. To provide some idea of what this means in practice, we implemented a brute-force search in Java which finds the optimal global bid for 100 discrete valuations and, in doing so, searches through 100 bids per valuation and per auction. Using these settings, on a $1.66GHz$ Intel Centrino the optimal solution is found within 4 seconds when $m = 10$, and takes about 35 seconds when $m = 15$.

When bidding in multiple auctions, we can show that it is optimal to bid higher in preferred auctions. Intuitively, this is because, in case of second-price auctions, preferred auctions give an agent a higher probability of winning the good at a lower price. This observation can reduce the computation required since only bid values which are higher than the bids in less preferred auctions need to be considered. In practice this usually reduces the number of solutions for each auction which satisfy expression (12) to a single one (in which case the computation of the optimal global bid becomes linear in the number of auctions). In addition to the computational benefit, this observation also provides some guidelines and intuition about the strategies. The relationship between preferred auctions and the optimal global bid is more precisely described as follows:

**Theorem 8.** *Under Assumption 1, for any $i, j \in M, i \neq j$: if auction $i$ is (weakly) preferred over auction $j$, i.e., if $G_i(b) \geq G_j(b)$ for $0 \leq b \leq v_{max}$, and if auction $i$ is strictly preferred at both $b_i^*$ and $b_j^*$, where $b_i^*$ and $b_j^*$ maximise Equation (10), then $b_i^* > b_j^*$ for any $v > 0$.*

PROOF We first prove $b_i^* \geq b_j^*$ by contradiction, and then go on to show that $b_i^* \neq b_j^*$. Suppose the opposite holds and there exist $i, j \in M$ such that $b_i^*, b_j^* \in \mathbf{b}$ are optimal and $b_i^* < b_j^*$. We then show that the expected utility is strictly higher when interchanging the two bids, i.e., when bidding $b_i^*$ in auction $j$ and $b_j^*$ in auction $i$. Let $\mathbf{b}_{b_i^* \leftrightarrow b_j^*}$ denote the global bid $\mathbf{b}$ where the two bids $b_i^*, b_j^*$ are interchanged. We now show that $U(\mathbf{b}_{b_i^* \leftrightarrow b_j^*}, v) - U(\mathbf{b}, v)$ is strictly positive for $0 \leq v \leq v_{max}$. Using Equation (10) and integration by parts we obtain the following:

$$U(\mathbf{b}_{b_i^* \leftrightarrow b_j^*}, v) - U(\mathbf{b}, v) = (v\,c - b_i^*)(G_j(b_i^*) - G_i(b_i^*)) + (v\,c - b_j^*)(G_i(b_j^*) - G_j(b_j^*))$$

$$- v\,c\,G_j(b_i^*)G_i(b_j^*) + v\,c\,G_j(b_j^*)G_i(b_i^*) + \int_{b_i^*}^{b_j^*} G_i(y) - G_j(y)dy, \quad (13)$$

where $c = \prod_{k \in M \setminus \{i,j\}} (1 - G_k(b_k))$. First we note that $c > 0$ because the strict preference of the auctions at $b_i^*$ and $b_j^*$ requires that $0 < b_i^* < v_{max}, 0 < b_j^* < v_{max}$ (since $G_i(0) = G_j(0) = G_i(v_{max}) = G_j(v_{max}) = 0$), and therefore due to Theorem 7 it holds that $0 < b_k < v_{max}$ for $k \in M \setminus \{i, j\}$. Next, note that the last term in Equation (13) is always positive since $b_j^* > b_i^*$ and $G_i(b) \geq G_j(b)$. We can therefore ignore this term if the remaining part is also positive. Let this term be denoted by $\epsilon$ in the following. Since we assumed $b_i^*$ and $b_j^*$ to be optimal, from Equation (11) the following holds:

$$b_i^* = v\,c\,(1 - G_j(b_j^*)) \Leftrightarrow v\,c - b_i^* = v\,c\,G_j(b_j)),$$
$$b_j^* = v\,c\,(1 - G_i(b_i^*)) \Leftrightarrow v\,c - b_j^* = v\,c\,G_i(b_i)). \quad (14)$$

By replacing $v\,c - b_i^*$ and $v\,c - b_j^*$ in Equation (13) by $v\,c\,G_j(b_j)$ and $v\,c\,G_i(b_i)$ respectively, and by rearranging the terms, we obtain the following:

$$U(\mathbf{b}_{b_i^* \leftrightarrow b_j^*}, v) - U(\mathbf{b}, v) = v\,c\,(G_i(b_i^*) - G_j(b_i^*))(G_i(b_j^*) - G_j(b_j^*)) + \epsilon. \quad (15)$$

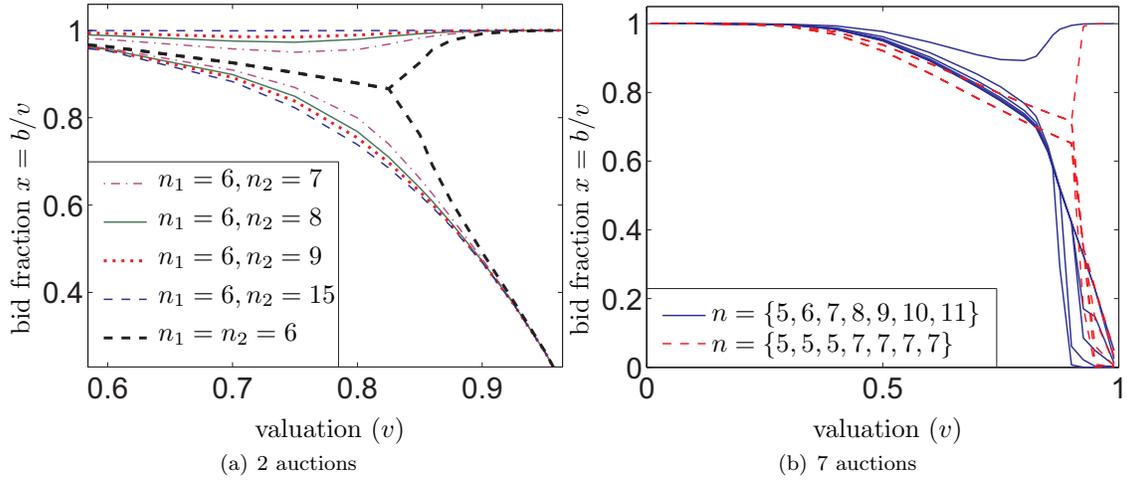

(a) 2 auctions          (b) 7 auctions

Figure 4: Optimal global bid as a fraction of the valuation for (a) two and (b) seven simultaneous auctions for various settings. The auctions differ in the number of local bidders which are present at each auction.

Now, since $G_i(b_i^*) > G_j(b_i^*)$, $G_i(b_j^*) > G_j(b_j^*)$, $v > 0$, $c > 0$, and $\epsilon \geq 0$, it follows that $U(\mathbf{b}_{b_i^* \leftrightarrow b_j^*}, v) > U(\mathbf{b}, v)$. As a result, $b_i^*$ and $b_j^*$ cannot be optimal bids for auctions $i$ and $j$ respectively, and this contradicts our initial proposition.

The above proves that $b_i^* \geq b_j^*$. We now show that $b_i^*$ must be *strictly* higher than $b_j^*$. This follows directly from Equation (14). Suppose $b_i^* = b_j^*$, then from Equation (14) it follows that $G_i(b_i^*) = G_j(b_i^*)$ and $G_i(b_j^*) = G_j(b_j^*)$. However, this conflicts with the requirement that auction $i$ must be strictly preferred at $b_i^*$ and $b_j^*$, thereby completing the proof.                                                                    □

The above proof is based on the requirement that auction $i$ is strictly preferred over auction $j$ for at least the two optimal bids in these auctions. Although this requires the optimal bid to be known in advance, the proof also applies to cases where auction $i$ is strictly preferred over auction $j$ over the entire range $0 < b < v_{max}$. This is the case, for example, when the number of bidders in auction $i$ is strictly less than in auction $j$. Furthermore, we note that similar results hold for slightly different conditions. Specifically, if the auction is strictly preferred at a set of points with non-zero measure anywhere within the range $[b_j^*, b_i^*]$, then $b_i^* \geq b_j^*$ since $\epsilon > 0$ in Equation (14). Consequently, the condition $b_i^* \geq b_j^*$ holds when auction $i$ is preferred over auction $j$ and $G_i, G_j$ can be described by *analytic functions*.

### 5.2.3 EMPIRICAL EVALUATION

In this section, we examine how the bidding strategy of the global bidder is affected when the auctions have different numbers of local bidders. To this end, we find the optimal bids that maximise the utility given by Equation (10) using a standard optimisation technique (Press, Flannery, Teukolsky, & Vetterling, 1992). The numerical results shown here assume the bidder valuation distribution to be uniform, although similar results are obtained with other commonly used distributions. Two different scenarios are evaluated. In the first, we investigate how the optimal bids change with increasing differences between the local bidder numbers in the case of two simultaneous auctions (see Figure 4(a)). Here, the number of local bidders in the first auction ($n_1$) is fixed to 6 whereas the number of local bidders in the second auction ($n_2$) is varied between 6 and 15 in the second (note that we depict the optimal strategy as a fraction of the valuation ($b_i/v$) rather than the actual bid value since this more clearly demonstrates the effect of preferred auctions). Figure 4(a) shows that the global bidder bids higher in the preferred auction (i.e., the one with fewer bidders), consistent with Lemma 8. As the valuation increases, the bid in the preferred auction first decreases relative to the valuation, and then increases, similar to the case with the same number of local bidders. However, the higher bid is now much closer to the true valuation, especially if the difference between the auctions is large.

The simulation is extended to more auctions in the second scenario. The results for two different settings are shown in Figure 4(b). In one setting (solid lines), each of the seven auctions has a different number of local bidders. In the other settings (dashed lines) the results are shown where 3 auctions have 5 local bidders each and the other 4 auctions have 7 local bidders. As before, we observe that the global bidder always bids higher in the auctions with fewer local bidders. Although the appearance of the bifurcation point is common for simultaneous auctions with the same number of bidders, this is not the case when the number of bidders is not identical. This is because the global bidder always bids different amounts for auctions with different numbers of bidders. As a result, the bifurcation phenomenon that indicates the transition from equal bids to high-low bids does not occur.

## 5.3 Sequential and Simultaneous Auctions

In this section we extend our analysis of the optimal bidding strategy to sequential auctions. Specifically, the auction process consists of $R$ rounds, and in each round any number of auctions are running simultaneously. Such a combination of sequential and simultaneous auctions is very common in practice, especially online.[15] It turns out that the analysis for the case of simultaneous auctions can be easily extended to include sequential auctions. In the following, the set of simultaneous auctions in round $r$ is denoted by $M^r$, and the vector of bids in that round by $\mathbf{b}^r$. As before, the analysis assumes that all other bidders are local and bid in a single auction. We initially assume that the global bidders have complete knowledge about the number of rounds and the number of auctions in each round but we then relax these assumptions.

The expected utility in round $r$, denoted by $U^r$, is similar to before (Equation (1) in Section 4.1) except that now additional benefit can be obtained from future auctions if the desired item is not won in one of the current set of simultaneous auctions. For convenience, $U^r(\mathbf{b}^r, M^r)$ is abbreviated to $U^r$ in the following. The expected utility thus becomes:

---

15. Rather than being purely sequential in nature, online auctions also often overlap (i.e., new auctions can start while others are still ongoing). In that case, however, it is optimal to wait and bid in the new auctions only after the outcome of the earlier auctions is known, thereby reducing the chance of unwittingly winning multiple items. Using this strategy, overlapping auctions effectively become sequential and can thus be analysed using the results in this section.

$$U^r = v \cdot P^r(\mathbf{b}^r) - \sum_{i \in M^r} \left( \int_0^{b_i^r} yg(y)dy \right) + U^{r+1} \cdot (1 - P^r(\mathbf{b}^r))$$

$$= U^{r+1} + (v - U^{r+1})P^r(\mathbf{b}^r) - \sum_{i \in M^r} \left( \int_0^{b_{ri}} yg(y)dy \right), \tag{16}$$

where $P^r(\mathbf{b}^r) = 1 - \prod_{i \in M^r}(1 - G(b_i^r))$ is the probability of winning at least one auction in round $r$. Now, we take the partial derivative of Equation (16) in order to find the optimal bid $b_j^r$ for auction $j$ in round $r$:

$$\frac{\partial U^r}{\partial b_j^r} = g(b_j^r)\left[(v - U^{r+1})\prod_{i \in M^r \setminus \{j\}}(1 - G(b_i^r)) - b_j^r\right]. \tag{17}$$

Note that Equation (17) is almost identical to Equation (4) in Section 4.3, except that the valuation $v$ is now replaced by $v - U^{r+1}$. The optimal bidding strategy can thus be found by backward induction (where $U^{R+1} = 0$) using the procedure outlined in Section 4.3.

Now, we first relax the assumption that the global bidder has complete knowledge about the number of auctions in *future* rounds. Let $p(m)$ denote the probability that there are $m$ auctions in the next round and let $m_{max}$ denote the maximum number of auctions. Furthermore, let $M_j$ be the set of $j$ auctions and $U^r(\mathbf{b}^r, M_j)$ the expected utility in round $r$ when there are $j$ auctions in that round. The uncertainty about the number of auctions can be incorporated into Equations (16) and (17) by replacing $U^{r+1}$ with $\sum_{j=0}^{m_{max}} p(j)U^{r+1}(\mathbf{b}^{r+1}, M_j)$. Furthermore, uncertainty about the number of rounds can be addressed by adding a discount factor $0 < \delta \leq 1$ which represents the probability that there are no more auctions after this round. Finally, we note that the equation can be similarly extended to non-identical auctions and to settings where $G$ depends on the number of auctions and/or the round.

## 6. Market Efficiency

Efficiency is an important system-wide property since it characterises to what extent the market maximises social welfare (i.e., the sum of utilities of all agents in the market). To this end, in this section we study the efficiency of markets with either static or dynamic local bidders, and the impact that a global bidder has on the efficiency in these markets. Specifically, efficiency in this context is maximised when the bidders with the $m$ highest valuations (recall that $m$ denotes the number of auctions, but also the total number of items) in the entire market obtain a single item each. For simplicity we assume that the total number of bidders in the market is at least $m$. More formally, we define the efficiency of an allocation as:

**Definition 1. Efficiency of Allocation.** *The efficiency $\eta_K$ of an allocation $K$ is the obtained social welfare proportional to the maximum social welfare that can be achieved in any type of market and is given by:*

$$\eta_K = \frac{\sum_{i=1}^{n_T} v_i(K)}{\sum_{i=1}^{n_T} v_i(K^*)}, \tag{18}$$

where $K^* = \arg\max_{K \in \mathcal{K}} \sum_{i=1}^{n_T} v_i(K)$ *is an efficient allocation,* $\mathcal{K}$ *is the set of all possible allocations (assuming bidders can be allocated to any auction),* $v_i(K)$ *is bidder* $i$'s utility for the allocation $K \in \mathcal{K}$, and $n_T$ *is the total number of bidders participating in the market.*

Now, in order to measure the efficiency of the market and the impact of a global bidder, we run simulations for the markets with and without a global bidder and for the different types of local bidders. The experiments are carried out as follows. Each bidder's valuation (both local and global) is independently drawn from a uniform distribution with support $[0, 1]$. In the experiments without a global bidder, an additional local bidder is placed in one of the auctions such that the overall number of bidders is the same on average compared to the case with a global bidder. The local bidders bid their true valuations, whereas the global bidder bids optimally in each auction as described in Section 4.3. The experiments are repeated 10000 times in order to get statistically significant results with a 99% confidence interval.

The results of these experiments are shown in Figure 5. Note that a degree of inefficiency is inherent to a multi-auction market with only local bidders.[16] For example, if there are two auctions selling one item each, and the two bidders with the highest valuations both bid locally in the same auction, then the bidder with the second-highest value does not obtain the good. Thus, the allocation of items to bidders is inefficient, and this inefficiency increases as the number of auctions increases (keeping the average number of bidders per auction equal). As can be observed from Figure 5, however, the efficiency increases when $n$ becomes larger. This is because the differences between the bidders with the highest valuations become smaller, thereby decreasing the loss of efficiency.

Furthermore, Figure 5 shows that if one of the local bidders is replaced by a global bidder this generally creates a positive effect on the efficiency when the number of bidders is small, but that no significant change occurs when there are many local bidders (this holds both for static and dynamic local bidders). The latter comes as no surprise since the impact of a single bidder diminishes as there are more bidders competing in an auction. The former, on the other hand, is not obvious; the introduction of a global bidder potentially leads to a decrease of efficiency since this bidder can unwittingly win more than one item. Furthermore, a global bidder will generally bid below its true valuation which can also result in inefficient outcomes. However, the results show that, on average, the opposite occurs. This is because, when there is no global bidder and only few local bidders, there is a high probability that a local bidder with a low valuation will win the item. A global bidder, on the other hand, is likely to win that auction if it has a sufficiently high valuation (even though the global bids are below the true valuation, as was shown in Section 4.3.4 bids are often uniform and fairly close to the true value). This effect is particularly pronounced in the case of dynamic local bidders since it may occur that an auction has no local bidder whatsoever, in which case the global bidder wins the item for sure.

---

16. An exception is when $n = 1$ and bidders are static, since the market is then completely efficient without a global bidder. However, since this is a very special case and does not apply to other settings, we do not discuss it further here.

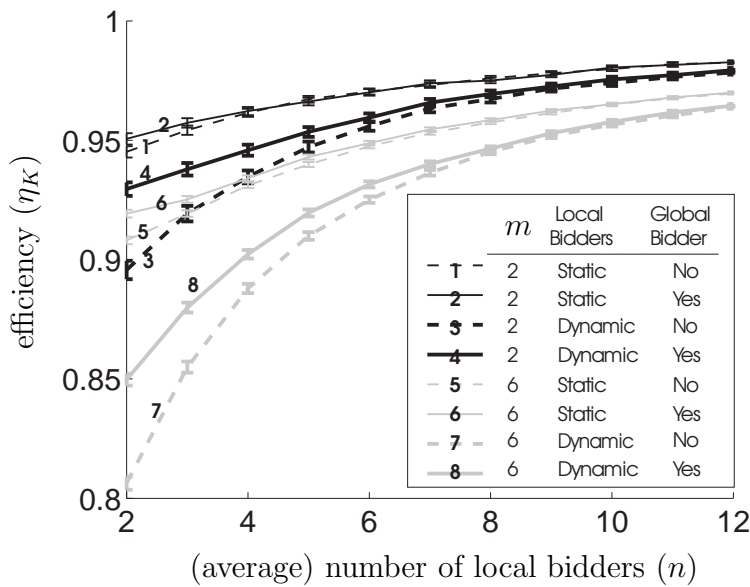

Figure 5: Average efficiency for different market settings as shown in the legend, where $m$ is the number of auctions (items), and $n$ is the *average* number of local bidders per auction. The error-bars indicate 99% confidence intervals.

## 7. Conclusions

In this paper, we derive utility-maximising strategies for an agent that has to bid in multiple, simultaneous second-price auctions. We first analyse the case where a single global bidder bids in all auctions, whereas all other bidders are local and bid in a single auction. For this setting, we find that it is optimal to place non-zero bids in all auctions that sell the desired item, even when a bidder requires only a single item and derives no additional benefit from having more. Thus, a potential buyer can achieve considerable benefit by participating in multiple auctions and employing an optimal bidding strategy.

For most common valuation distributions, we show analytically that the optimal bids for identical auctions consist of at most two values, a high bid and a low bid, and that it is optimal to bid high in at most one auction. Moreover, by writing the high bid in terms of the low ones, the problem of finding the optimal bids reduces to optimising a single variable. This considerably simplifies the original optimisation problem and can thus be used in practice to compute the optimal bids for any number of auctions. Furthermore, we show that, when the number of auctions becomes large, it is optimal to bid uniformly across all auctions. We also analyse the setting where auctions are not identical and differ in their probability of winning. Although it is still optimal to participate in all auctions, we find that the best strategy is then to bid relatively high in more favourable auctions (i.e., where the probability of winning is highest). We investigate other practical considerations as well. We show that budget constraints limit the number of auctions that bidders participate in. Specifically, if a global bidder's budget is equal to or less than its valuation, the optimal strategy reverts to bidding in a single auction under certain conditions, thereby justifying the presence of local bidders. Furthermore, we consider sequential auctions and show that our results can be readily applied to markets where auctions occur both sequentially and simultaneously. Finally, we compare the efficiency of a market with multiple simultaneous auctions with and without a global bidder. We show that, if a local bidder is replaced by a global one, this increases the average efficiency and thus social welfare of the sysstem when the number of bidders is small, but that no significant effect is found if there are many local bidders.

There are a number of interesting directions for future work. First of all, whereas this paper focuses on the buyer, we intend to extend the analysis and consider the revenue from the sellers' point of view. This is particularly relevant when sellers have a more pro-active role and can set auction parameters such as the reserve price to try and maximise their expected revenue. This is closely related to the research on competing sellers, where it is shown that the optimal auction parameters depend on the competition with other sellers, since this affects the number of potential buyers that are attracted to a particular auction (McAfee, 1993; Peters & Severinov, 1997; Burguet & Sákovics, 1999; Gerding et al., 2007b). However, the current literature on competing sellers assumes that the buyers only participate in a single auction, which is shown here to be suboptimal. The case where sellers optimise auction parameters and buyers can participate in any number of auctions simultaneously has so far not been investigated. Furthermore, this paper has taken a decision-theoretic approach by analysing the case of a single global bidder. An interesting open problem is characterising the game-theoretic solution in the case that multiple global bidders interact strategically.

## Acknowledgments

This paper has been significantly extended from a very preliminary version that was published previously (Gerding, Dash, Yuen, & Jennings, 2007a).

This research was undertaken as part of the EPSRC (Engineering and Physical Research Council) funded project on Market-Based Control (GR/T10664/01). This is a collaborative project involving the Universities of Birmingham, Liverpool and Southampton and BAE Systems, BT and HP. This research was also undertaken as part of the ALADDIN (Autonomous Learning Agents for Decentralised Data and Information Systems) project and is jointly funded by a BAE Systems and EPSRC strategic partnership. In addition, we would like to thank Alex Rogers, Ioannis Vetsikas, Wenming Bian, and Florin Constantin for their input. The authors are also very grateful to Adam Prügel-Bennett for his valuable help with some of the proofs, and Richard Engelbrecht-Wiggans for the useful discussions and comments. Finally, we would like to thank the reviewers for their thorough and detailed feedback.

## Appendix A. Proofs

### A.1 Reduction of Search Space

**Lemma 2.** *Under Assumption 1, if $\lambda_G(b)$ is non-decreasing in $(0, v_{max})$, then the function $H(b) = b(1 - G(b))$ has a unique critical point $b^f$ in that interval.*

Proof At a critical point of $H$ the following equation must hold:

$$\frac{d}{db}H(b) = \frac{d}{db}\left[b(1 - G(b))\right] = 1 - bg(b) - G(b)$$

$$= g(b)\left[\frac{1 - G(b)}{g(b)} - b\right] = g(b)\left[\frac{1}{\lambda_G(b)} - b\right] = 0. \tag{19}$$

Now, since $g(b) > 0$ (due to Assumption 1), equality (19) only holds when $1/\lambda_G(b) - b = 0$. Therefore, in order to show that (19) has at most one solution, it is sufficient to show that $1/\lambda_G(b) - b$ is either strictly increasing or strictly decreasing. However, since at the boundaries $b = 0$ and $b = v_{max}$ we have $h(0) = 1$ and $h(v_{max}) = -v_{max}g(v_{max})$, and since $g(b) > 0$, we only need to consider the latter. Since $-b$ is strictly decreasing, it is sufficient to show $1/\lambda_G(b)$ is non-increasing. By assumption we have that $\lambda_G(b)$ is non-decreasing, and thus $1/\lambda_G(b)$ is non-increasing. □

**Theorem 3.** *If $\lambda_F(b)$ is non-decreasing, then $\lambda_G(b)$ and $\lambda_{\widehat{G}}(b)$ are also non-decreasing, where $G(b) = F(b)^n$ and $\widehat{G} = e^{n(F(y)-1)}$ for $n \geq 2$.*

Proof In the case of static local bidders we prove that $\frac{1-G(b)}{g(b)}$ is a non-increasing function, thus showing that $\frac{g(b)}{1-G(b)}$ is a non-decreasing function. In more detail, we can refactor $\frac{1-G(b)}{g(b)}$ as follows:

$$\frac{1 - G(b)}{g(b)} = \frac{1 - F^n(b)}{nF^{n-1}(b)f(b)}$$

$$= \left[\frac{1 - F(b)}{f(b)}\right]\left[\frac{1 + F(b) + \ldots + F^{n-1}(b)}{nF^{n-1}(b)}\right] \tag{20}$$

$$= \frac{1}{n}\left[\frac{1 - F(b)}{f(b)}\right]\left[1 + F^{-1}(b) + \ldots + F^{1-n}(b)\right].$$

Due to the non-decreasing hazard rate of $f(b)$, we can derive that $\frac{1-F(b)}{f(b)}$ is non-increasing as follows:

$$\frac{\partial}{\partial b}\left(\frac{1 - F(b)}{f(b)}\right) = -\frac{\partial}{\partial v}\left(\frac{f(b)}{1 - F(b)}\right)\left[\frac{1 - F(b)}{f(b)}\right]^2 \leq 0$$

since $\left[\frac{1-F(b)}{f(b)}\right]^2 \geq 0$. Furthermore, the first order condition on the second part of equation 20, $[1 + F^{-1}(b) + \ldots + F^{1-n}(b)]$ is $-F^{-2}(b) - \ldots - (n-1)F^{-n}(b)$, is negative for all $b$. Thus, this second part is strictly decreasing. Therefore, given that the first part is non-increasing and the second part is strictly decreasing, it implies that overall $\frac{1-G(b)}{g(b)}$ is strictly decreasing. Hence $\lambda_G(b)$ is non-decreasing.

In the case of dynamic local bidders, we can rewrite $\lambda_{\widehat{G}}(b)$ as:

$$
\begin{aligned}
\frac{\hat{g}(b)}{1 - \widehat{G}(b)} &= \frac{nf(b)e^{n(F(b)-1)}}{1 - e^{n(F(b)-1)}} \\
&= \left[\frac{nf(b)}{1 - F(b)}\right]\left[\frac{(1 - F(b))e^{n(F(b)-1)}}{1 - e^{n(F(b)-1)}}\right] \\
&= \left[\frac{nf(b)}{1 - F(b)}\right]\left[\frac{1 - F(b)}{e^{n(1-F(b))} - 1}\right].
\end{aligned}
\tag{21}
$$

Since $\lambda_F$ is non-decreasing, clearly the first part of equation (21) is non-decreasing. We shall now prove that the second part of equation (21), $\gamma(b) = \frac{1-F(b)}{e^{n(1-F(b))}-1}$, is also non-decreasing. The first order condition on $\gamma(b)$ is given by:

$$
\frac{\delta\gamma(b)}{\delta b} = \frac{-f(b)(e^{n(1-F(b))} - 1) + (1 - F(b))(nf(b)e^{n(1-F(b))})}{(e^{n(1-F(b))} - 1)^2}.
\tag{22}
$$

Since the denominator of equation (22) is always non-negative, it is thus sufficient to show that the numerator is always non-negative. The numerator of equation (22) can be rewritten as:

$$
f(b)e^{n(1-F(b))}\left[n - 1 - nF(b) + e^{n(F(b)-1)}\right],
$$

in which the first and second terms are $> 0$ $\forall b$. Thus, in order to prove that the numerator is always non-negative, it remains to be shown that the third term, $\left[n-1-nF(b)+e^{n(F(b)-1)}\right]$, is non-negative. This term is equal to $n - 1 + e^{-n} > 0$ (since $n > 1$) and 0 at the extremums $b_{min}$ and $b_{max}$ respectively. Furthermore, the first order differential on this term yields $-nf(b)[1 - e^{n(F(b)-1)}] < 0$, $\forall b$. Hence the third term is also $> 0$, $\forall b$. □

**Lemma 3.** *If $H(b) = b(1 - G(b))$ has a unique critical point then*

$$
b > b^f \qquad \Longleftrightarrow \qquad g(b) > (1 - G(b))/b,
$$

*and similarly*

$$
b < b^f \qquad \Longleftrightarrow \qquad g(b) < (1 - G(b))/b.
$$

PROOF As we saw already, if $H$ has a unique critical point then $b > b^f$ iff $H$ is decreasing at $b$. Therefore

$$
b > b^f \iff H'(b) < 0 \iff 1 - G(b) - bg(b) < 0 \iff g(b) > (1 - G(b))/b.
$$

The other result follows analagously. □

**Theorem 4.** *Under Assumptions 1 and 2, the global bid $\mathbf{b}^*$ that maximises Equation 1 has at most one high bid, i.e., at most one $i \in M$ for which $b_i > b^f$, where $b^f$ is the unique critical point of $H(b) = b(1 - G(b))$.*

Proof The second derivatives of $U$ at an interior critical point $\mathbf{b}$ are as follows:

$$
\begin{aligned}
\frac{\partial^2 U}{\partial b_i^2} &= g'(b_i)\Big(v\prod_{j\neq i}(1-G(b_j))-b_i\Big)-g(b_i) = -g(b_i), \\
\frac{\partial^2 U}{\partial b_i \partial b_j} &= -g(b_i)g(b_j)v\prod_{k\neq i,j}(1-G(b_k)) = -g(b_i)\frac{g(b_j)b_j}{1-G(b_i)}.
\end{aligned}
\tag{23}
$$

The proof is by contradiction: we show that if there are two high bids then the critical point cannot be a local maximum of $U$. This is done by showing that if there are two high bids at $\mathbf{b}$, then the Hessian matrix with entries (23) has a positive eigenvalue, or equivalently that there exists a vector $\mathbf{a} = (a_1, a_2, \ldots, a_m)$ for which

$$
\sum_{i,j} a_i a_j \frac{\partial^2 U}{\partial b_i \partial b_j}(\mathbf{b}) > 0.
\tag{24}
$$

If we can show that the Hessian matrix at $\mathbf{b}$ has a positive eigenvalue, this means that $\mathbf{b}$ is either a local minimum or a saddle point (Magnus & Neudecker, 1999, Chapter 6), which in turn means that a small enough displacement in the direction of $\mathbf{a}$ leads to an increase in $U$, contradicting the local optimality of $\mathbf{b}$.

Assume without loss of generality that the auctions are rearranged such that $b_1 = b_2 = b_+ > b^f$. From Lemma 3 this implies that $g(b_+) > (1-G(b_+))/b_+$. Then we choose $\mathbf{a} = (1, -1, 0, \ldots, 0)$, so that (24) becomes:

$$
\begin{aligned}
\sum_{i,j} a_i a_j \frac{\partial^2 U}{\partial b_i \partial b_j}(\mathbf{b}) &= \frac{\partial^2 U}{\partial b_1^2}(\mathbf{b}) + \frac{\partial^2 U}{\partial b_2}(\mathbf{b}) - 2\frac{\partial^2 U}{\partial b_1 \partial b_2}(\mathbf{b}) \\
&= 2g(b_+)\left(\frac{g(b_+)b_+}{1-G(b_+)}-1\right) \\
&> 0.
\end{aligned}
$$

$\square$

## A.2 Limit Cases

In this section we examine a global bidder's expected utility when the number of auctions goes to infinity. In particular, we prove that for a large enough number of auctions, the optimal behaviour is always to bid uniformly.

First of all, note from (5) and the definition of the optimal uniform bid $\mathbf{b}_m^{u*}$, that

$$
\mathbf{b}_m^{u*} = v(1-G(\mathbf{b}_m^{u*}))^{m-1}.
\tag{25}
$$

**Lemma 4.** *The smallest bid $b_-$ in the optimal (possibly non-uniform) bid vector $\mathbf{b}$ tends to 0 uniformly in $v$ as $m \to \infty$.*

Proof This follows from the fact that the expected payment function $EP(b) = \int_0^b x g(x)\,dx$ is a strictly monotonically increasing continuous function of $b$, and hence has a monotonically increasing continuous inverse, $EP^{-1} : [0, EP(v_{max})] \rightarrow [0, v_{max}]$ with $EP^{-1}(0) = 0$.

$$
\begin{aligned}
0 < U(\mathbf{b}) \quad &< \quad v - mEP(b_-) \\
\implies \quad b_- \quad &< \quad EP^{-1}(v/m) \\
&< \quad EP^{-1}(v_{max}/m) \\
&\rightarrow \quad 0 \quad as \; m \rightarrow \infty
\end{aligned}
$$

Our first step towards proving that uniform bidding is eventually globally optimal is to show that the utility from optimal bidding converges to its maximum value of $v$:

**Theorem 5.** *Under Assumption 1, the expected utility as defined by Equation 1 from playing the optimal uniform bid $\mathbf{b}_m^{u*}$ converges to $v$, in the sense that for all $\epsilon > 0$ there is a constant $m_\epsilon$ such that $m > m_\epsilon$ implies $U(\mathbf{b}_m^{u*}) > v - \epsilon$.*

PROOF First of all note that the result is trivial for $v < \epsilon$ since $U(\mathbf{b}_m^{u*}) \geq 0$ always. From the definition of utility (1), and using (25),

$$
\begin{aligned}
U(\mathbf{b}_m^{u*}) \quad &= \quad v\Big(1 - (1 - G(\mathbf{b}_m^{u*}))^m\Big) - mEP(\mathbf{b}_m^{u*}) \\
&= \quad v - \mathbf{b}_m^{u*}(1 - G(\mathbf{b}_m^{u*})) - mEP(\mathbf{b}_m^{u*}). \tag{26}
\end{aligned}
$$

From Lemma 4 we know that $\mathbf{b}_m^{u*} \rightarrow 0$, which implies $\mathbf{b}_m^{u*}(1 - G(\mathbf{b}_m^{u*})) \rightarrow 0$, so we turn our attention to the expected payment term, showing that it tends to zero as $m \rightarrow \infty$. From the definition (25) of $\mathbf{b}_m^{u*}$,

$$
m = \frac{\ln(b_m^*/v)}{\ln(1 - G(b_m^*))} + 1,
$$

so that

$$
\begin{aligned}
\lim_{m \rightarrow \infty} mEP(\mathbf{b}_m^{u*}) \quad &= \quad \lim_{m \rightarrow \infty} \left( \frac{\ln(\mathbf{b}_m^{u*}/v)}{\ln(1 - G(\mathbf{b}_m^{u*}))} + 1 \right) EP(\mathbf{b}_m^{u*}) \\
&= \quad \lim_{m \rightarrow \infty} \frac{\ln(\mathbf{b}_m^{u*}/v)EP(\mathbf{b}_m^{u*})}{\ln(1 - G(\mathbf{b}_m^{u*}))}. \tag{27}
\end{aligned}
$$

To prove that the above limit is zero we demonstrate the stronger limit

$$
\lim_{b \rightarrow 0} \frac{\ln(b/v)EP(b)}{\ln(1 - G(b))} = 0, \tag{28}
$$

which we do by applying L'Hopital's rule multiple times. First we apply it to the numerator of (28):

$$
\begin{aligned}
\lim_{b \rightarrow 0} \Big( \ln(b/v)EP(b) \Big) \quad &= \quad \lim_{b \rightarrow 0} \frac{EP(b)}{(\ln(b/v))^{-1}} \\
&= \quad \lim_{b \rightarrow 0} \frac{bg(b)}{b^{-1}(\ln(b/v))^{-2}} \\
&= \quad \lim_{b \rightarrow 0} \big(bg(b)\big)\big(b(\ln(b/v))^2\big) \\
&= \quad 0,
\end{aligned}
$$

where the last limit holds uniformly on $\epsilon \leq v \leq v_{max}$.

Now that the numerator has been shown to converge to zero, it is possible to apply L'Hopital's rule directly to (28),

$$
\begin{aligned}
\lim_{b \to 0} \frac{\ln(b/v)EP(b)}{\ln(1 - G(b))} &= \lim_{b \to 0} \frac{b^{-1}EP(b) + g(b)b\ln(b/v)}{-g(b)(1 - G(b))^{-1}} \\
&= -\lim_{b \to 0} \left( \frac{EP(b)}{bg(b)} + b\ln(b/v) \right) \quad (29) \\
&= -\lim_{b \to 0} \left( \frac{bg(b)}{g(b) + bg'(b)} + b\ln(b/v) \right). \quad (30)
\end{aligned}
$$

$g'(b) \geq 0$ for all small enough $b$ in order for $g(b) \geq 0$ to be true everywhere, which implies that

$$
\frac{bg(b)}{g(b) + bg'(b)} < b \to 0,
$$

and the limit $b\ln(b/v) \to 0$ is obvious, so the limit in (30) must be zero. This in turn implies that the expected payment from uniform bidding tends to zero, while the expected value tends to $v$, and thus the theorem is proved. □

## A.3 Budget Constraints

In what follows we provide the proof for the proposition that the optimal strategy is to bid in a single auction under certain conditions, if the exposure is constrained to at most its valuation. The proof for any number of auctions is complex. Therefore, we first provide a formal proof for the case of two auctions, and then generalise the result to more than two. The proof for the two-auction case also provides a building block for the inductive proofs contained in the more general case.

First we show that the unconstrained optimum bid eventually exceeds any given budget:

**Theorem 6.** *Under Assumption 1 and assuming that $g(b)$ is bounded throughout $[0, v_{max}]$, for all $v > 0$ and for any $C$, there exists an $m$ for which the exposure of the optimal global bid $\mathbf{b}^*$ maximising Equation (1) exceeds $C$, i.e., $\sum_{i \in M} b_i > C$.*

PROOF To begin with, note that the conditions of Corollary 1 are met, so that in fact we can restrict attention to uniform bidding when $m$ is sufficiently large, in particular as $m \to \infty$. What we then show is that for all $v > V > 0$ and any $C > 0$ there is an $m$ such that $m\mathbf{b}_m^{u*} > C$.

The proof is by contradiction. Assume that there is a constant $C$ such that $\mathbf{b}_m^{u*}m < C$ for all $m$. By the bounded density assumption, $G(\mathbf{b}_m^{u*}) \leq K\mathbf{b}_m^{u*} < KC/m$, so that (using Equation (25)):

$$
\begin{aligned}
\mathbf{b}_m^{u*} &= v(1 - G(\mathbf{b}_m^{u*}))^{m-1} \\
&> v(1 - KC/m)^{m-1} \\
&> V(1 - KC/m)^{m-1} \\
&\to Ve^{-KC} \quad \text{as } m \to \infty. \quad (31)
\end{aligned}
$$

The fact that $\mathbf{b}_m^{u*}$ is thus bounded below contradicts the bound $\mathbf{b}_m^{u*}m < C$. $\qquad\square$

The following Lemma characterises the bidding strategy when the budget is $C \leq v$ and $m = 2$. This result is then used in Theorem 7 to prove the more general case where $m \geq 2$.

**Lemma 5.** *For $m = 2$ the optimal bidding strategy of a global bidder with budget $C \leq v$ is given by $\mathbf{b}^{c*} = (C, 0)$ when the probability density function $g(x)$ is convex and $g(0) = 0$.*

PROOF In order to prove the lemma, we consider the difference between the optimal bid $\mathbf{b}^{c*}$ and an arbitrary bid $\mathbf{b}' = b_1, b_2$ where $\mathbf{b}' \neq \mathbf{b}^{c*}$ and $b_1 + b_2 = C$:

$$U(\mathbf{b}^{c*}, v) - U(\mathbf{b}', v) = vG(C) - \int_0^C yg(y)dy - \Big[ v\big[1 - (1 - G(b_1))(1 - G(b_2))\big]$$
$$- \int_0^{b_1} yg(y)dy - \int_0^{b_2} yg(y)dy \Big].$$

Hence,

$$\begin{aligned}
U(\mathbf{b}^{c*}, v) - U(\mathbf{b}', v) &= (v - C)G(C) + \int_0^C G(y)dy - vG(b_1) - vG(b_2) + b_1 G(b_1) \\
&\quad + vG(b_1)G(b_2) - \int_0^{b_1} G(y)dy + b_2 G(b_2) - \int_0^{b_2} G(y)dy \\
&= (v - C)(G(C) - G(b_1) - G(b_2)) + \int_{b_2}^C G(y)dy \\
&\quad - (C - b_1)G(b_1) - (C - b_2)G(b_2) - \int_0^{b_1} G(y)dy \\
&\quad + vG(b_1)(G(b_2) \\
&= (v - C)(G(C) - G(b_1) - G(b_2)) + \int_{b_2}^C G(y)dy - b_2 G(b_1) \\
&\quad - b_1 G(b_2) - \int_0^{b_1} G(y)dy + vG(b_1)G(b_2).
\end{aligned} \tag{32}$$

Now, to prove that $\mathbf{b}^{c*}$ is indeed the optimal bid, we need to show that the above difference is positive for any $\mathbf{b}'$ that satisfies the budget constraint. In order to do so, we first separate the above equation into two parts, namely, $X(b_1, b_2) = \int_{b_2}^C G(y)dy - b_2 G(b_1) - b_1 G(b_2) - \int_0^{b_1} G(y)dy$ and $(v - C)Y$ where $Y(b_1, b_2) = (G(C) - G(b_1) - G(b_2))$. We now show that both $X$ and $Y$ are positive; thus implying that, since $(v - C)$ is positive, $U(\mathbf{b}^{c*}, v) - U(\mathbf{b}', v)$ is also positive.

Now, we rewrite $X$ in terms of $b1$ only by replacing $b2$ with $C - b_1$, and $X$ becomes:

$$X(b_1) = \int_{C-b_1}^C G(y)dy - (C - b_1)G(b_1) - b_1 G(C - b_1) - \int_0^{b_1} G(y)dy. \tag{33}$$

In order to find the local maxima and minima we set the derivative $\frac{\partial X}{\partial b_1}$ to zero:

$$\begin{aligned}
\frac{\partial X}{\partial b_1} &= G(C - b_1) - (C - b_1)g(b_1) + G(b_1) + b_1 g(C - b_1) - G(C - b_1) - G(b_1) \\
&= b_1 g(C - b_1) - (C - b_1)g(b_1) = 0.
\end{aligned} \tag{34}$$

Since $g(0) = 0$, it is easy to see that there are *at least* three solutions to Equation (34), namely $b_1 = \{0, C, C/2\}$. We shall now show that the first derivative of $X$ is always non-negative within the range $b_1 = [0, C/2]$, implying that $X$ increases within this range. Since $X(0) = 0$ and the solution is symmetric around $C/2$, it follows that $X$ is always non-negative. More formally, we need to show that:

$$\frac{\partial X}{\partial b_1} \geq 0, b_1 \in [0, C/2].$$

We now prove that this holds if $g(x)$ is convex. Now, from the definition of a convex function we have:

$$g(\lambda x_1 + (1 - \lambda)x_2)) \leq \lambda g(x_1) + (1 - \lambda)g(x_2), \lambda \leq 1.$$

Now, let $x_2 = 0$. Then:

$$g(\lambda x_1) \leq \lambda g(x_1) + (1 - \lambda)g(0).$$

Since $g(0) = 0$, this becomes:

$$\begin{aligned}
g(\lambda x_1) &\leq \lambda g(x_1) \Rightarrow \\
\frac{1}{\lambda}g(\lambda x_1) &\leq g(x_1).
\end{aligned} \tag{35}$$

Let $x_1 = (C - b_1)$ and $\lambda = \frac{b_1}{C - b_1}$. Then, Equation (35) becomes:

$$\begin{aligned}
\frac{C - b_1}{b_1}g(b_1) &\leq g(C - b_1) \Rightarrow \\
b_1 g(C - b_1) - (C - b_1)g(b_1) &\geq 0.
\end{aligned} \tag{36}$$

Thus, this shows that $\frac{\partial X}{\partial b_1} \geq 0$, which, in turn, proves that $X \geq 0$.

We now prove that $Y = G(C) - G(b_1) - G(b_2) \geq 0$. Since we assume that $g(x)$ is convex and positive in the interval $[0, 1]$ and that $g(0) = 0$, it follows that $g(x)$ is increasing and thus $G(x) = \int_0^x g(x)dx$ is also convex. As a result, we can use the condition in Equation (35) for $G(x)$. Now, by replacing $\lambda$ by $\frac{x}{b_1 + b_2}$ and $x_1$ by $b_1 + b_2$, we have $G(x) \leq \frac{G(b_1 + b_2)}{b_1 + b_2}x$ for any $x \leq b_1 + b_2$ and it follows that:

$$\begin{aligned}
G(b_1) + G(b_2) &\leq b_1 \frac{G(b_1 + b_2)}{b_1 + b_2} + b_2 \frac{G(b_1 + b_2)}{b_1 + b_2} \Rightarrow \\
G(b_1) + G(b_2) &\leq G(b_1 + b_2)\left(\frac{b_1}{b_1 + b_2} + \frac{b_2}{b_1 + b_2}\right) \Rightarrow \\
G(b_1) + G(b_2) &\leq G(b_1 + b_2).
\end{aligned}$$

Hence, $Y = G(C) - G(b_1) - G(b_2) \geq 0$. As a result, $U(\mathbf{b}^*, v) - U(\mathbf{b}', v) \geq 0$ for any $b_1 + b_2 = C$. Note that so far we have assumed that the sum of bids is equal to the budget constraint, and we have not mentioned the case where $b_1 + b_2 < C$. We now show that it is optimal to bid the full budget (i.e., $b_1 + b_2 = C$).

Consider any arbitrary $B' = b_1, b_2$ where $b_1 + b_2 = S \leq C$. Then, by replacing $C$ with $S$ in the above, we know that $U((S, 0), v) \geq U((b_1, b_2), v)$. Hence, it remains to be shown that $U((C, 0), v) \geq U((S, 0), v)$. We again consider the difference between these two bids, which from Equation (1) is:

$$
\begin{aligned}
U((C, 0), v) - U((S, 0), v) &= vG(C) - \int_0^C yg(y)dy - vG(S) + \int_0^S yg(y)dy \\
&= v\left(G(C) - G(S)\right) - \left(CG(C) - \int_0^C G(y)dy\right) \\
&\quad + \left(SG(S) - \int_0^S G(y)dy\right) \\
&= (v - C)(G(C) - G(S)) + \int_S^C G(y)dy \\
&\quad - (C - S)G(S).
\end{aligned}
$$

Now, $G(C) - G(S)$ is positive since $S \leq C$ by definition and $G(x)$ is non decreasing and, therefore, $\int_S^C G(y)dy - (C - S)G(S)$ is also positive. Hence, $U((C, 0), v) - U((S, 0), v) \geq 0$. $\square$

We have thus shown that a global bidder will bid in only one auction if its budget is constrained such that $C \leq v$ when bidding in two simultaneous auctions, provided that $g(x)$ is convex and $g(0) = 0$ This results can now be generalised to the case that $m \geq 2$. In more detail, the following theorem holds:

**Theorem 7.** *Under Assumptions 1 and 3, if the global bidder has a budget $C \leq v$, then Equation (7) is satisfied for $\mathbf{b}^{c*} = (C, 0, \ldots, 0)$, i.e., it is optimal to bid $C$ in exactly one auction.*

PROOF As before, we consider the difference between the optimal bid $\mathbf{b}^{c*}$ and another bid $\mathbf{b}' = (b_1, \ldots, b_m)$ where $\mathbf{b}' \neq \mathbf{b}^{c*}$ and $C = \sum_{i \in M} b_i$, and show that this is positive:

$$
\begin{aligned}
U(\mathbf{b}^{c*}, v) - U(\mathbf{b}', v) &= vG(C) - \int_0^C yg(y)dy - \left(v\left[1 - \prod_{i \in M}(1 - G(b_i))\right]\right. \\
&\quad \left. - \sum_{i \in M}\int_0^{b_i} yg(y)dy\right) \\
&= (v - C)G(C) + \int_0^C G(y)dy - v\left[1 - \prod_{i \in M}(1 - G(b_i))\right] \\
&\quad + \sum_{i \in M}\left(b_iG(b_i) - \int_0^{b_i} G(y)dy\right).
\end{aligned}
\tag{37}
$$

By inserting $(v - C) \sum_{i \in M} G(b_i) - (v - C) \sum_{i \in M} G(b_i)$ into Equation (37) this can be rewritten as:

$$U(\mathbf{b}^{c*}, v) - U(\mathbf{b}', v) = (v - C) \left( G(C) - \sum_{i \in M} G(b_i) \right) + \int_0^C G(y) dy$$

$$- C \sum_{i \in M} G(b_i) - \sum_{i \in M} \left[ \int_0^{b_i} G(y) dy - b_i G(b_i) \right]$$

$$- v \left[ 1 - \prod_{i \in M} (1 - G(b_i)) - \sum_{i \in M} G(b_i) \right]$$

$$= (v - C) \left( G(C) - \sum_{i \in M} G(b_i) \right) + \int_0^C G(y) dy$$

$$- \sum_{i \in M} \left( (C - b_i) G(b_i) + \int_0^{b_i} G(y) dy \right)$$

$$- v \left[ 1 - \prod_{i \in M} (1 - G(b_i)) - \sum_{i \in M} G(b_i) \right].$$

Now, define the variables $X_m$ and $Z_m$ as:

$$X_m = \int_0^C G(y) dy - \sum_{i \in M} \left( (C - b_i) G(b_i) + \int_0^{b_i} G(y) dy \right),$$

$$Z_m = - \left[ 1 - \prod_{i \in M} (1 - G(b_i)) - \sum_{i \in M} \right].$$

Then, we can rewrite Equation (37) as:

$$U(\mathbf{b}^{c*}, v) - U(\mathbf{b}', v) = (v - C) \left( G(C) - \sum_{i \in M} G(b_i) \right) + X_m + v Z_m. \tag{38}$$

We now prove that each part of the equation is positive (i.e., $(v-C)(G(C) - \sum_{i \in M} G(b_i)) \geq 0, X_m \geq 0, vZ \geq 0$). We first prove that $G(C) - \sum_{i \in M} G(b_i) \geq 0$ using the relationship in Equation (35) for convex functions. Specifically, by taking $\lambda = \frac{b_i}{C}$ and since $C = \sum_{i \in M} b_i$ this yields:

$$G(C) \geq \frac{C}{b_i} G(b_i) \Rightarrow$$

$$\sum_{i \in M} b_i G(C) \geq \sum_{i \in M} C G(b_i) \Rightarrow$$

$$\frac{\sum_{i \in M} b_i}{C} G(C) \geq \sum_{i \in M} G(b_i) \Rightarrow$$

$$G(C) \geq \sum_{i \in M} G(b_i).$$

Since, by definition of the case being studied, $v \geq C$, we have $(v-C)(G(C)-\sum_{i \in M} G(b_i)) \geq 0$.

We now prove that $X_m \geq 0$ using an inductive argument. In more detail, let $S = \sum_{i \in M \setminus \{m\}} b_i = C - b_m$. Then $X_m$ can be written in terms of $X_{m-1}$ as follows:

$$X_m = \int_0^C G(y)dy - \sum_{i \in M} \left( (C-b_i)G(b_i) + \int_0^{b_i} G(y)dy \right)$$

$$= \int_0^S G(y)dy - \sum_{i \in M \setminus \{m\}} \left( (S-b_i)G(b_i) + \int_0^{b_i} G(y)dy \right) + \int_S^C G(y)dy -$$

$$(C-b_m)G(b_m) - (C-S) \sum_{i \in M \setminus \{m\}} G(b_i) - \int_0^{b_m} G(y)dy$$

$$= X_{m-1} + \int_S^C G(y)dy - (C-b_m)G(b_m) - b_m \sum_{i \in M \setminus \{m\}} G(b_i) - \int_0^{b_m} G(y)dy.$$

Since $G(S) = G(C - b_m) \geq \sum_{i \in M \setminus \{m\}} G(b_i)$, then :

$$X_m \geq X_{m-1} + \int_{C-b_m}^C G(y)dy - (C-b_m)G(b_m) - b_m G(C-b_m) - \int_0^{b_m} G(y)dy.$$

From Lemma 5, we have shown that $\int_{C-b_m}^C G(y)dy - (C-b_m)G(b_m) - b_m G(C-b_m) - \int_0^{b_m} G(y)dy \geq 0$ and therefore $X_m \geq X_{m-1}$. The base case is $X_2$ which has been shown within the proof of Lemma 5 to be positive. Hence $X_m \geq 0$.

We again use an inductive argument to finally prove that $Z_m \geq 0$. The base case of $Z_m$ is when $m = 2$ which yields:

$$Z_2 = -\Big[ 1 - \prod_{i \in M} (1 - G(b_i)) - \sum_{i \in M} G(b_i) \Big]$$

$$= -\Big[ 1 - (1 - G(b_1))(1 - G(b_2)) - G(b_1) - G(b_2) \Big]$$

$$= G(b_1)G(b_2)$$

$$\geq 0.$$

The inductive hypothesis is then formulated as $Z_m \geq 0$ if $Z_{m-1} \geq 0$. In order to prove this hypothesis, we express $Z_m$ as a function of $Z_{m-1}$:

$$Z_m = \prod_{i \in M} (1 - G(b_i)) + \sum_{i \in M} G(b_i) - 1$$

$$= (1 - G(b_m)) \prod_{i \in M \setminus \{m\}} (1 - G(b_i)) + \sum_{i \in M} G(b_i) - 1$$

$$= (1 - G(b_m)) \left( Z_{m-1} - \sum_{i \in M \setminus \{m\}} G(b_i) + 1 \right) + \sum_{i \in M} G(b_i) - 1$$

$$= Z_{m-1} - G(b_m) Z_{m-1} + G(b_m) \sum_{i \in M \setminus \{m\}} G(b_i)$$

$$= Z_{m-1} + G(b_m) \left( \sum_{i \in M \setminus \{m\}} G(b_i) - Z_{m-1} \right)$$

$$= Z_{m-1} + G(b_m) \left( 1 - \prod_{i \in M \setminus \{m\}} (1 - G(b_i)) \right)$$

$$\geq Z_{m-1}.$$

The above thus proves the inductive step, which along with the base case, thereby proves that $Z_m \geq 0$. Hence the third part of Equation (38) is also positive. Since all three parts of this equation are positive, this implies that $U(\mathbf{b}^{c*}, v)$ is indeed optimal when $\mathbf{b}^{c*} = (C, 0, \ldots, 0)$. Note that we assumed $C = \sum_{i \in M} b_i$. However, using the same argument as in Lemma 5 it is easy to see that it is indeed optimal to bid the full budget. $\qquad \square$

**Corollary 2.** *Under Assumption 1, if either $g(0) > 0$ or $g(x)$ is strictly concave, and the global bidder has a budget $C = v$, then Equation (7) is satisfied by bidding strictly positive in at least two auctions.*

PROOF The proof is largely based on the reverse arguments from Lemma 5. Without loss of generality we take $m = 2$. Let $\mathbf{b}^L = \{C, 0\}$ denote the single-auction bid and $\mathbf{b}' = \{b_1, b_2\}$ an arbitrary bid such that $\mathbf{b}' \neq \mathbf{b}^L$ and $b_1 + b_2 = C$, $b_1, b_2 \geq 0$ as before. Since $v - C = 0$ we have $U(\mathbf{b}^L, v) - U(\mathbf{b}', v) = X(b_1)$, where X is given by Equation (33). Furthermore, $\frac{\partial X}{\partial b_1} = b_1 g(C - b_1) - (C - b_1) g(b_1)$ (see Equation (34)). Now, in order to prove that it is optimal to bid strictly positive in both auctions, it is sufficient to show that there exists a $C > b_1 > 0$ such that $X(b_1) < 0$ when one of the two conditions holds.

We first show that this holds when $g(0) > 0$. It is easy to see that $\frac{\partial X}{\partial b_1}(0) = -Cg(0) < 0$ and $X(0)$ is thus strictly decreasing. Since $X(0) = 0$ this proves that $X(b_1) < 0$ for $b_1$ slightly larger than zero. Since $b_2 = C - b_1$ it follows that $b_2 > 0$ as long as $b_1 < C$, and an agent can thus do better by bidding in both auctions.

We now consider the case that $g(0) = 0$ but $g(x)$ is *strictly* concave. By replacing the condition for convex functions in Lemma 5 with that of strictly concave functions it follows that $\frac{1}{\lambda} g(\lambda x_1) > g(x_1)$ for $\frac{1}{2} C > x_1 > 0$. By setting $x_1 = (C - b_1)$ and $\lambda = \frac{b_1}{C - b_1}$ as before, this gives (see Equations (35) and (36)) $\frac{\partial X}{\partial b_1} = b_1 g(C - b_1) - (C - b_1)g(b_1) < 0$ for $b_1 \in (0, C/2)$. $\qquad\square$